\documentclass[fleqn,usenatbib]{mnras}

\usepackage{mathptmx}
\usepackage[T1]{fontenc}
\usepackage{ae,aecompl}
\usepackage{graphicx}	% Including figure files
\usepackage{amsmath}	% Advanced maths commands
\usepackage{amssymb}	% Extra maths symbols
\usepackage{verbatim} % Comment environment 
\usepackage[english]{babel}
\usepackage[dvipsnames]{xcolor}
\usepackage{natbib}
\usepackage{txfonts}
\usepackage{multirow}
\usepackage{pifont}

\title[He enrichment in intermediate-age clusters]{Helium enrichment in intermediate-age Magellanic Clouds clusters: towards an ubiquity of multiple stellar populations?}

\author[W. Chantereau et al.]{
W. Chantereau,$^{1}$\thanks{E-mail: w.chantereau@ljmu.ac.uk}
M. Salaris,$^{1}$
N. Bastian,$^{1}$
S. Martocchia$^{2,1}$
\\
$^{1}$Astrophysics Research Institute, Liverpool John Moores University, 146 Brownlow Hill, Liverpool L3 5RF, UK \\
$^{2}$European Southern Observatory, Karl-Schwarzschild-Stra{\ss}e 2, D-85748 Garching bei M{\"u}nchen, Germany
}

\date{Accepted XXX. Received YYY; in original form ZZZ}

\pubyear{2018}

\begin{document}

\maketitle

\begin{abstract}
  Intermediate-age star clusters in the Magellanic Clouds harbour signatures of the multiple stellar populations long thought to be restricted to old
  globular clusters. We compare synthetic horizontal branch models with \textit{Hubble Space Telescope} photometry of clusters in the Magellanic Clouds, with age between $\sim$2 and $\sim$10~Gyr,
  namely NGC~121, Lindsay~1, NGC~339, NGC~416, Lindsay~38, Lindsay~113, Hodge~6 and NGC~1978. We find a clear signature of initial helium abundance spreads ($\Delta Y$)
  in four out of these eight clusters (NGC~121, Lindsay~1, NGC~339, NGC~416) and we quantify the value of $\Delta Y$. For two clusters
  (Lindsay~38, Lindsay~113) we can only determine an upper limit for $\Delta Y$, whilst for the two youngest clusters in our sample (Hodge~6 and NGC~1978) no conclusion
  about the existence of an initial He spread can be reached.
  Our $\Delta Y$ estimates are consistent with the correlation between maximum He abundance spread and mass of the host cluster found in Galactic globular clusters.
  This result strengthens the emerging view that the formation of
  multiple stellar populations is a standard process in massive star clusters, not limited to a high redshift environment.  
\end{abstract}

        
%%%%%%%%%%%%%%%%%%%%%%%%%%%%%%%%%%%%%%%%%%%%%%%%%%
%%%%%%%%%%%%%%%%% BODY OF PAPER %%%%%%%%%%%%%%%%%%        
%%%%%%%%%%%%%%%%%%%%%%%%%%%%%%%%%%%%%%%%%%%%%%%%%%

\section{Introduction}\label{introduction}

The multiple stellar populations (MPs) present in individual globular clusters (GCs) are characterised by star-to-star abundance anti-correlations 
of light elements (C-N, O-Na and Mg-Al to a certain extent) together with spreads of initial He abundances \citep[e.g.][]{Milone18,Bastian18}. 
It has been recently shown that massive intermediate-age clusters in the Magellanic Clouds (MCs) --with ages down to $\sim$2~Gyr--
also display light element abundance patterns like GCs 
\citep[e.g.][]{Hollyhead17,Niederhofer17_121,Niederhofer17}. On the other hand, clusters younger than $\sim$2~Gyr seem to lack detectable MPs, 
suggesting that age (or stellar mass) play a major factor in the onset of this phenomenon in massive stellar clusters \citep{Martocchia18}. 

An important question to be addressed is the following: Do the MCs massive clusters older than $\sim$2~Gyr also display He abundance spreads, like Galactic GCs?  
If this is the case, these intermediate-age clusters are the counterparts of Galactic GCs in terms of MPs, thus suggesting that the MP formation  
is not restricted to high redshift environments. This, in turn, implies that young stellar clusters can also be used to constrain the MP formation process. 

In a very recent paper, \citet{Lagioia19} determined the presence of He abundance spread in four SMC massive clusters, employing photometry of
red giant branch (RGB) stars.
They found small helium abundance spreads in NGC~121, NGC~339 and NGC~416, while no spread was found for Lindsay~1.

Here, we will investigate the presence of a He abundance spread in a sample of MC clusters by modelling 
the morphology of their Red Clump (RC) and red horizontal branch (HB) stars in the colour-magnitude-diagram (CMD) using synthetic HB (and RC) models.
As is well known, the CMD morphology of the He-burning phase is very sensitive to the initial He abundance of the parent populations, 
and indeed synthetic HB models have been employed to determine He abundance spreads 
in Milky Way GCs such as NGC~104 \citep{Gratton13}, NGC~2419 \citep{diCriscienzo11,diCriscienzo15}, 
NGC~2808 \citep{Dalessandro11}, NGC~5272 \citep{Dalessandro13}, NGC~5904 \citep{Gratton13}, NGC~6388 \citep{Busso07}, and NGC~6441 \citep{Busso07,Caloi07}. 

The massive, intermediate-age clusters investigated in this study are Lindsay~1, NGC~121, NGC~339, NGC~416, in common with \citet{Lagioia19}, plus
Hodge~6,  Lindsay~38, Lindsay~113 and NGC~1978. They are all younger than the average Milky Way GC, with ages ranging between $\sim$2~Gyr and $\sim$10~Gyr. Additionally, all clusters have had MP signatures detected within them either photometrically or spectroscopically, except Lindsay 38 and Lindsay 113 that are currently being investigated (Martocchia et al.~2019, in preparation).

Our study expands the sample of clusters in the MCs investigated for the presence of initial He abundance spreads. Also, our method is complementary to the technique employed
by \citet{Lagioia19}. These latter authors model several colour differences --sensitive to He, C, N, O abundance spreads-- between fiducial sequences that
trace the RGB of the main populations of each cluster (for one cluster they also determine the He spread from the RGB bump, whose brightness is also
sensitive to the initial He abundance). As such, their method tends to measure differences of mean He abundances between cluster subpopulations.  
Our HB modelling aims at reproducing the full colour and magnitude range of the observed HBs, and should estimate the maximum He spread amongst stars in individual clusters. 

The paper is organised as follows. 
Sect.~\ref{mando} that describes both stellar evolution models and observations employed in this paper.   
Section~\ref{theory} describes briefly the synthetic HB models, how they can reveal the presence of initial $Y$ variations, and the fitting
procedure to observational data.
In Sect.~\ref{analysis}  
we investigate the presence of $Y$ variations in individual clusters in our sample, and in Sect.~\ref{discussion} we finally discuss and summarise our results.

\section{Stellar models and observations}\label{mando}

We employ non-rotating stellar evolution models and tracks computed with the code STAREVOL \citep[e.g.][]{Lagarde12}. Our calculations do not include atomic 
diffusion\footnote{All observed HBs and RCs investigated here are cool enough ($T_{\rm eff} \lesssim$10'000~K) to avoid strong effects of atomic diffusion \citep{HuiBonHoa00,Michaud11}.}. 
For each assumed cluster metallicity and age, we have computed models --from the zero age main sequence to the end of the HB, following the evolution 
through the He-flash-- with various values of the initial Helium mass fraction ($Y$), choosing appropriate initial main sequence masses to reach the cluster age at the beginning of the 
He-burning phase. Our calculations do not include the early-asymptotic giant branch phase following the exhaustion of central He.
The $Y$ values range from the value expected from Galactic chemical evolution ($\Delta Y$/$\Delta Z \sim 1.57$) to the maximum values given in Table~\ref{Table:ClustersHB}, 
that vary from cluster to cluster.

As for the metal distribution of our models, we assume a scaled solar distribution (\citealt{Asplund09}, with
an $\alpha$-enhancement for the case of NGC~104 and NGC~121, see next section). Also, the He-enhanced models 
(that in principle should have metal distributions with altered C, N, O, Na, Mg, Al abundances) are calculated for the same scaled solar (or $\alpha$-enhanced) metal mixture, 
given that stellar evolution is not affected by these abundance variations if the sum of the C+N+O abundance is kept constant at fixed 
metallicity \citep[as generally observed, within the errors, in Galactic GCs, see e.g.][]{Yong15}. In addition, we work on CMDs in the \textit{ACS} and \textit{WFC3}  
F475W, F555W, and F814W photometric bands, that are insensitive to variations of these light elements \citep[see e.g.,][]{Salaris06,Sbordone11}.  

Mass-loss during the red giant branch (RGB) evolution is accounted for by employing the Reimers formula \citep{Reimers75}:
\begin{equation*}
\dot{M} = 4 \times10^{-13} \eta_R \frac{LR}{M} M_\odot yr^{-1} 
\end{equation*}
where L, M and R are the model luminosity, mass and radius in solar units.
For each metallicity and $Y$ abundances we have calculated tracks for various values of $\eta_R$.

Bolometric corrections to the \textit{ACS} and \textit{WFC3} filters are obtained by interpolation amongst the tables from the MIST database \citep{Choi16} \footnote{\url{http://waps.cfa.harvard.edu/MIST/model_grids.html}}.

\begin{figure*}
   \centering
  \includegraphics[width=0.33\textwidth]{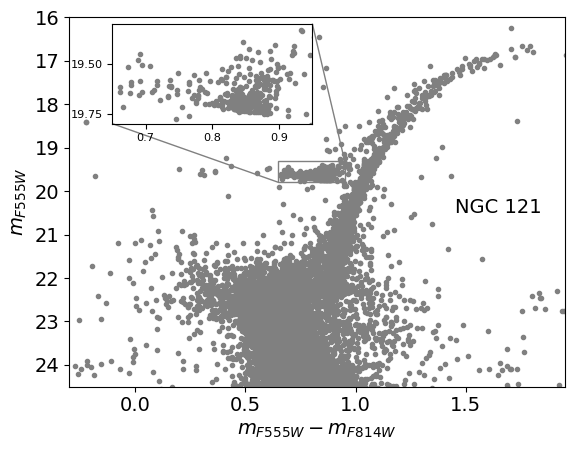}  
  \includegraphics[width=0.33\textwidth]{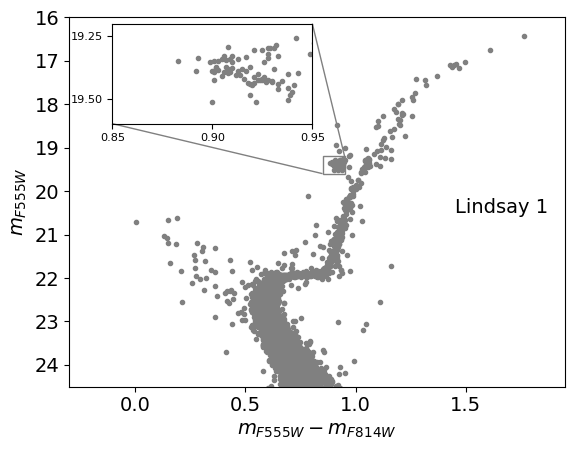}   
  \includegraphics[width=0.33\textwidth]{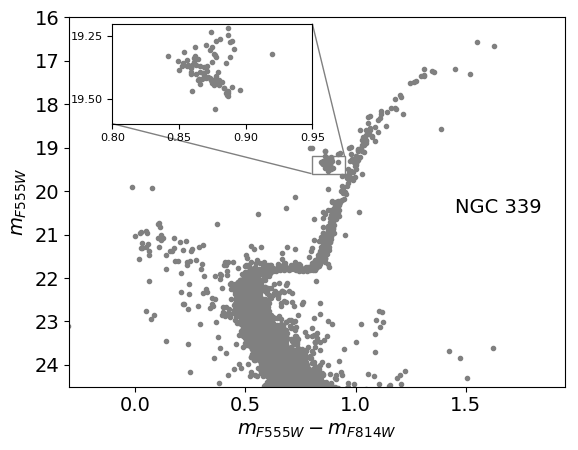} \\   
  \includegraphics[width=0.33\textwidth]{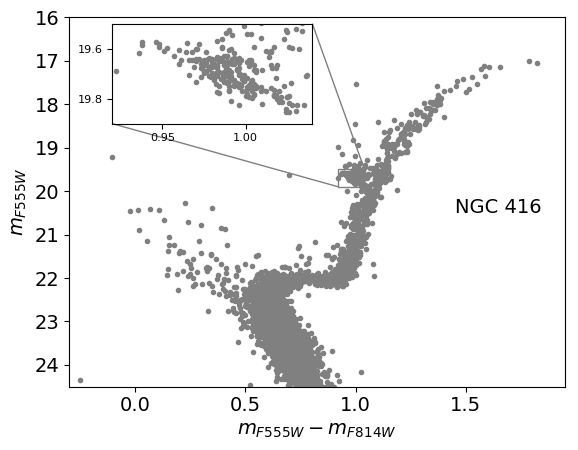} 
  \includegraphics[width=0.33\textwidth]{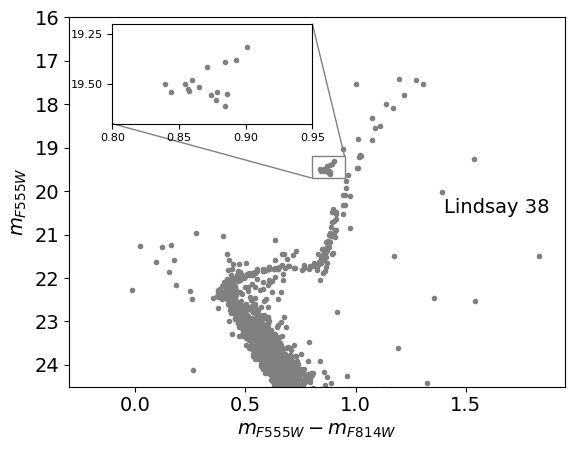}    
  \includegraphics[width=0.33\textwidth]{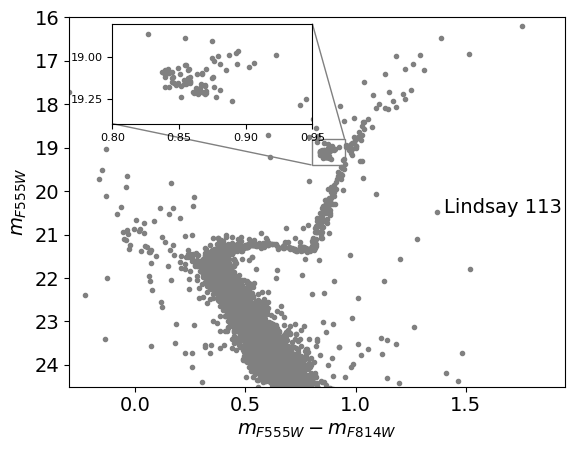} \\
  \includegraphics[width=0.33\textwidth]{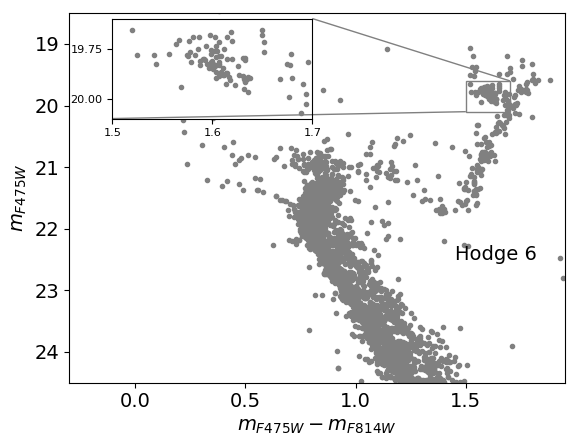} 
  \includegraphics[width=0.33\textwidth]{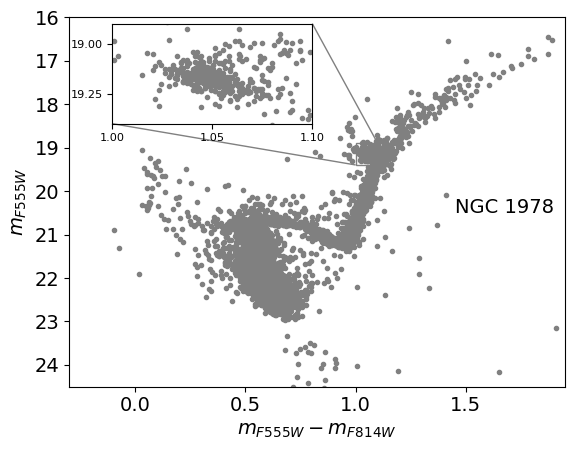} \\  
    \caption{CMDs of NGC~121, Lindsay~1, NGC~339, NGC~416, Lindsay~38, Lindsay~113, Hodge~6 and NGC~1978 with a zoom on the HB/RC region. 
Except for Hodge~6, whose CMD is displayed in the F475W vs (F475W-F814W) diagram, all the other CMDs are F555W vs (F555W-F814W).} \label{Figure:CMDs}
\end{figure*}

Clusters' photometries are taken from the \textit{Hubble Space Telescope} survey presented in \cite{Niederhofer17_121,Niederhofer17,Martocchia18} 
and Martocchia et al. (\textit{in prep.}). In this study we use the \textit{ACS} $F555W$ and $F814W$ optical filters, except 
for Hodge~6, for which we use \textit{WFC3} photometry in the $F475W$ and $F814W$ filters. 
The cluster CMDs are shown in Fig.~\ref{Figure:CMDs}, and the relevant cluster properties are listed in Table~\ref{Table:ClustersHB}. 

\cite{Niederhofer17_121,Niederhofer17,Martocchia18} and Martocchia et al. (\textit{in prep.}) investigated these clusters for differential reddening and only NGC~416 is affected (we refer to these works for more details). Thus we use the data corrected for differential reddening for this cluster.

\begin{table*}
\centering
\begin{tabular}{| c | c | c | c | c | c | c | c | c | c | c | c | c |}
	\hline \hline
    \multicolumn{9} {c} {Cluster} & \multicolumn{2} {c} {minimum $Y$ models} & \multicolumn{2} {c} {maximum $Y$ models} \\
    ID & [Fe/H] & Age (Gyr) & Ref. &  Mass (M$_\odot$) & Ref. &$(m-M)_V$ &  $E(B-V)$ & Ref. & Y$_\mathrm{ini}$ & M$_\mathrm{ini}$ (M$_\odot$) & Y$_\mathrm{ini}$ & M$_\mathrm{ini}$ (M$_\odot$) \\ \hline 
    NGC~104 & -0.72 & 12.0 & M15 & $7.79\times10^5$ & B18 & 13.37 & 0.04 & H96 & 0.251 & 0.905 & 0.291 & 0.84 \\ \hline   
	NGC~121 & -1.30 & 10.5 & G8a,N17 & $5.83\times10^5$ & G11 & 19.00 & 0.03 & G8a,N17a & 0.248 & 0.89 & 0.288 & 0.83 \\
    Lindsay~1 & -1.14 & 7.5$\pm$0.5 & G8b & $1.74\times10^5$ & G11 & 18.78 & 0.02 & G8b & 0.249 & 0.97 & 0.279 & 0.92 \\	
    NGC~339 & -1.12 & 6$\pm$0.5 & G8b & $2.88\times10^5$ & G11 & 18.80 & 0.02 & G8b & 0.250 & 1.04 & 0.290 & 0.97 \\	
    NGC~416 & -1.00 & 6$\pm$0.5 & G8b & $2.32\times10^5$ & G11 & 18.90 & 0.08 & G8b & 0.250 & 1.045 & 0.330 & 0.905 \\ 
    Lindsay~38 & -1.50 & 6$\pm$0.5 & M19 & $3.35\times10^4$ & G11 & 19.10 & 0.02 & M19 & 0.249 & 1.02 & 0.269 & 0.985 \\ 
    Lindsay~113 & -1.40 & 4.5$\pm$0.5 & M19 & $\sim2.3\times10^4$ & C10 & 18.85 & 0.02 & M19 & 0.249 & 1.11 & 0.269 & 1.07 \\      
    Hodge~6 & -0.40 & 2.25$\pm$0.05 & P14,G14 & $5.5\times10^4$ & G14 & 18.77 & 0.09 & P14 & 0.258 & 1.53 & 0.318 & 1.37\\ 
    NGC~1978 & -0.35 & 1.9$\pm$0.1 & M07 & $2-4\times10^5$ & W97 & 18.71 & 0.05 & M18b & 0.258 & 1.60 & 0.288 & 1.51 \\
    \hline \hline
\end{tabular}
\caption[]{Adopted parameters for the clusters we investigate in this study. Note that the minimum and maximum $Y$ models displayed in this table are the  models of our grid we use to interpolate in between to create the synthetic HB models (cf text). References: (M15) \cite{McDonald15}; (B18) \cite{Baumgardt18}; (H96) \cite{Harris96}, 2010 edition; (G8a) \cite{Glatt08_121}; (N17) \cite{Niederhofer17_121}; (G11) \cite{Glatt11}; (G8b) \cite{Glatt08}; (M19) Martocchia et al. (in prep.); (C10) Computed from the absolute magnitude in the V band \citep[-5.29,][]{Carretta10} and adopting a mass-to-light ratio of $\sim2$ \citep{Baumgardt18}; (P14) \cite{Piatti14}; (G14) \cite{Goudfrooij14}; (M07) \cite{Mucciarelli07}; (W97) \cite{Westerlund97}, (M18b) \cite{Martocchia18_1978}.} \label{Table:ClustersHB}
\end{table*}

\section{Synthetic horizontal branch modelling}\label{theory}

To determine the theoretical cluster HB (or RC) location and morphology in the CMD we need to fix a number of parameters, namely the cluster 
age, metallicity, initial He distribution, RGB mass loss efficiency ($\eta_R$, that determines the actual mass of the synthetic HB stars for a given cluster age and initial chemical composition). 
For each cluster, we fix age and metallicity to the values estimated in previous studies, as reported in Table~\ref{Table:ClustersHB}. 
Notice that variations of the age around the values in Table~\ref{Table:ClustersHB} will change the derived value of $\eta_R$ (because of a different HB progenitor mass) but not the 
overall results about the presence (or absence) of a He abundance spread in individual clusters.
Also, the minimum value of $Y$ (that we denote as the He abundance of the He-normal population) is fixed to the value given by $Y = Y_0+ \Delta Y/ \Delta Z \times Z$ where $Z$ is the heavy element mass fraction. The primordial helium mass fraction Y$_0$ chosen is equal to 0.2479 \citep{Coc04}.

The free parameters that are left to be determined by fitting synthetic HBs to observed CMDs are the minimum and eventually maximum value of $\eta_R$
(if the observed HB is matched with a spread of 
mass loss instead of $Y$), and the maximum value of $Y$ (if a range of $Y$ is required). 
For simplicity, we assume a uniform probability distribution for $\eta_R$ and $Y$, between the minimum and maximum values.   
We interpolate in $Y$ and $\eta_R$ amongst our model grid to determine the HB track of our synthetic star for a given $\eta_R$ and $Y$. We then extract a random age
with uniform probability between the 
zero age HB and the exhaustion of central He points, to fix the position of the synthetic stars in the CMD\footnote{The underlying standard assumptions is that stars are fed to the HB at a constant rate.}.
Magnitudes and colours of the synthetic stars are then perturbed by random Gaussian photometric errors, with 1$\sigma$ values taken from the mean photometric errors of the observations. We also checked these errors by comparing with the RGB width. We verticalised the RGB to determine the standard deviation of the $\delta$(colour) distribution of RGB stars at the HB magnitude level. The standard deviation then derived is similar to the photometric errors of the observations, in addition this standard deviation can be considered as an upper limit since the He spread also affects the RGB width. Thus we are confident with these photometric errors. For each cluster we create the same number of synthetic stars as the one observed in a box delimiting the HB region of each cluster. 

The procedure adopted to match the observed HB of a given cluster works as follows. For any given cluster we apply to the models distance modulus 
and reddening values listed in Table~\ref{Table:ClustersHB}, using the extinction coefficient for the \textit{ACS} and \textit{WFC3} filters from \cite{Goudfrooij09,Goudfrooij14}.
We then adjust $E(B-V)$ to fit the cluster RGB with the track of the HB progenitor, and fix $\eta_R$ to match the reddest part of the observed HB 
with models calculated with the minimum value of $Y$.
We then vary the maximum value of $Y$ at fixed $\eta_R$  --or $\eta_R$ and fixed initial $Y$-- to reproduce by eye the slope and full colour 
extension of the HB. Due to the strong sensitivity of the HB morphology to variations of $Y$ (and $\eta_R$), we found with numerical tests that 
a simple fit by eye can give an accuracy on $\Delta Y$ better than 0.01 (see Sect.~\ref{analysis}). 

We do not try to enforce the constraint of statistical agreement between the theoretical and observed star counts, because a perfect fit of star counts rests on the precise knowledge of, for example, the initial $Y$ distribution among the cluster stars, that could be extremely complicated and/or discontinuous.
The morphological constraints imposed on the matching synthetic HB are however sufficient to put strong bounds on $\Delta Y$, the maximum He abundance range, that is 
the main parameter discussed in this work. Obviously, our technique does not determine the exact number distribution of HB stars as a function of their initial $Y$.

\begin{figure*}
   \centering
  \includegraphics[width=0.45\textwidth]{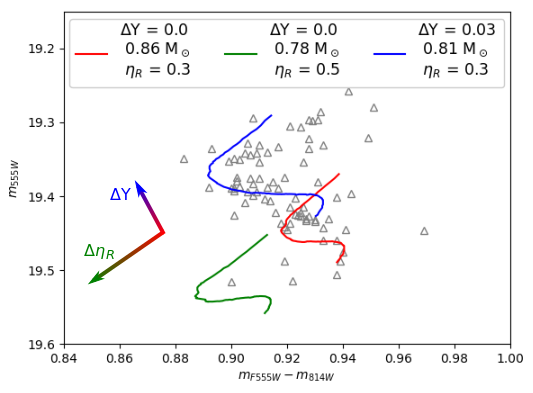} 
  \includegraphics[width=0.45\textwidth]{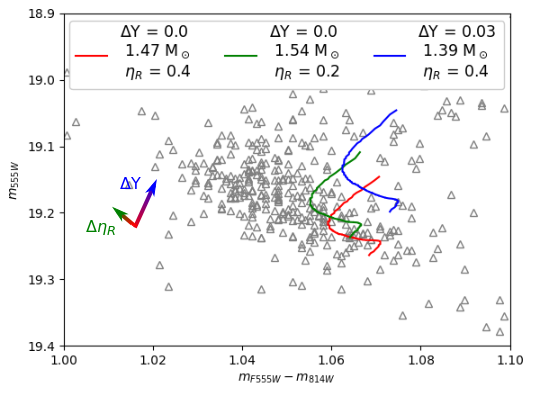}   
    \caption{\textbf{\textit{Left:}} HB of Lindsay 1 in grey open triangles. The HB stellar evolution tracks with initial main sequence mass 
$M_{ini} =$ 0.97~M$_\odot$, [Fe/H] = -1.14, $\Delta Y = 0$ (cluster age $\sim$7.5~Gyr) and $\eta_R$ = 0.3 and 0.5 are displayed with red and green lines, respectively. The track with initial main sequence mass  $M_{ini} =$ 0.92~M$_\odot$, [Fe/H] = -1.14, $\Delta Y = 0.03$ 
(HB age $\sim$7.5~Gyr), $\eta_R$ = 0.3 is displayed with a blue line. The values of the corresponding current HB masses are displayed in the labels. \textbf{\textit{Right:}} Horizontal branch of NGC~1978 
in grey open triangles. HB tracks with initial mass  $M_{ini} =$ 1.60~M$_\odot$, [Fe/H] = -0.35, $\Delta Y = 0$ (cluster age $\sim$1.9~Gyr), $\eta_R$ = 0.4 and 0.2 are  
displayed with red and green lines respectively. Tracks with an initial mass  $M_{ini} =$ 1.51~M$_\odot$, [Fe/H] = -0.35 and $\Delta Y = 0.03$ (HB age $\sim$1.9~Gyr) and $\eta_R$ = 0.4 is displayed with a 
blue line. The current masses are displayed in the label.} \label{Figure:Theory}
\end{figure*}

Figure~\ref{Figure:Theory} shows the case of two clusters, one (Lindsay~1) representative of intermediate-age and old clusters (initial mass of He-normal HB progenitors
lower than $\sim$1.5 $M_{\odot}$), 
and one (NGC~1978) representative of younger clusters but still populated by RGB stars with electron degenerate cores. 
For the sake of clarity we display selected HB evolutionary tracks without photometric errors applied. The tracks shown do not represent the best fit
models for these two clusters that will then
be presented in Sect.~\ref{analysis}, rather their purpose is just to highlight trends in the CMD.

In both cases a variation $\Delta Y$ at fixed $\eta_R$ (and age) moves the HB tracks in an orthogonal direction with respect to the effect of varying  
$\eta_R$ ($\Delta \eta_R$ ) at fixed $Y$ \citep[see also Fig.~1 in][]{Salaris16}, although the directions of the $\Delta Y$ and $\Delta \eta_R$ vectors change between the two age regimes. 
It is quite obvious even from this simple qualitative test shown in Fig.~\ref{Figure:Theory}, that Lindsay~1 HB morphology can be matched only 
with $\Delta Y>$0.
On the other hand, the HB morphology of NGC~1978 seems more likely to be shaped by a range of $\eta_R$.
We will see that the inclusion of photometric errors  
makes however difficult to draw firm conclusions for this cluster and the similar cluster Hodge~6.

We conclude this section with a test of our synthetic HB modelling on the well studied Galactic GC  NGC~104 
(total mass equal to $\sim$ $7.8\times10^{5} M_\odot$, age $\sim$12~Gyr, [Fe/H] = $-$0.72, as summarized in Table~\ref{Table:ClustersHB}) and compare with the synthetic HB modelling by \citet{Salaris16}, 
who found that a helium range $\Delta Y$=0.03 is needed to reproduce the observed HB morphology. Their result is in good agreement with several previous studies 
\citep{Anderson09, diCriscienzo10,Milone12,Gratton13} who determined $\Delta Y \sim$0.02-0.03 for this GC. 

\begin{figure*}
   \centering
  \includegraphics[width=0.4\textwidth]{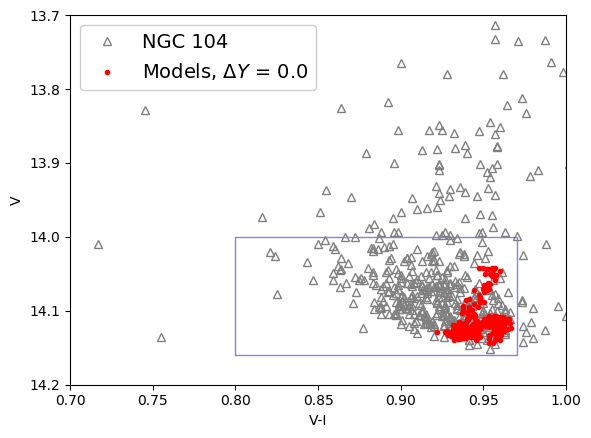}  
  \includegraphics[width=0.4\textwidth]{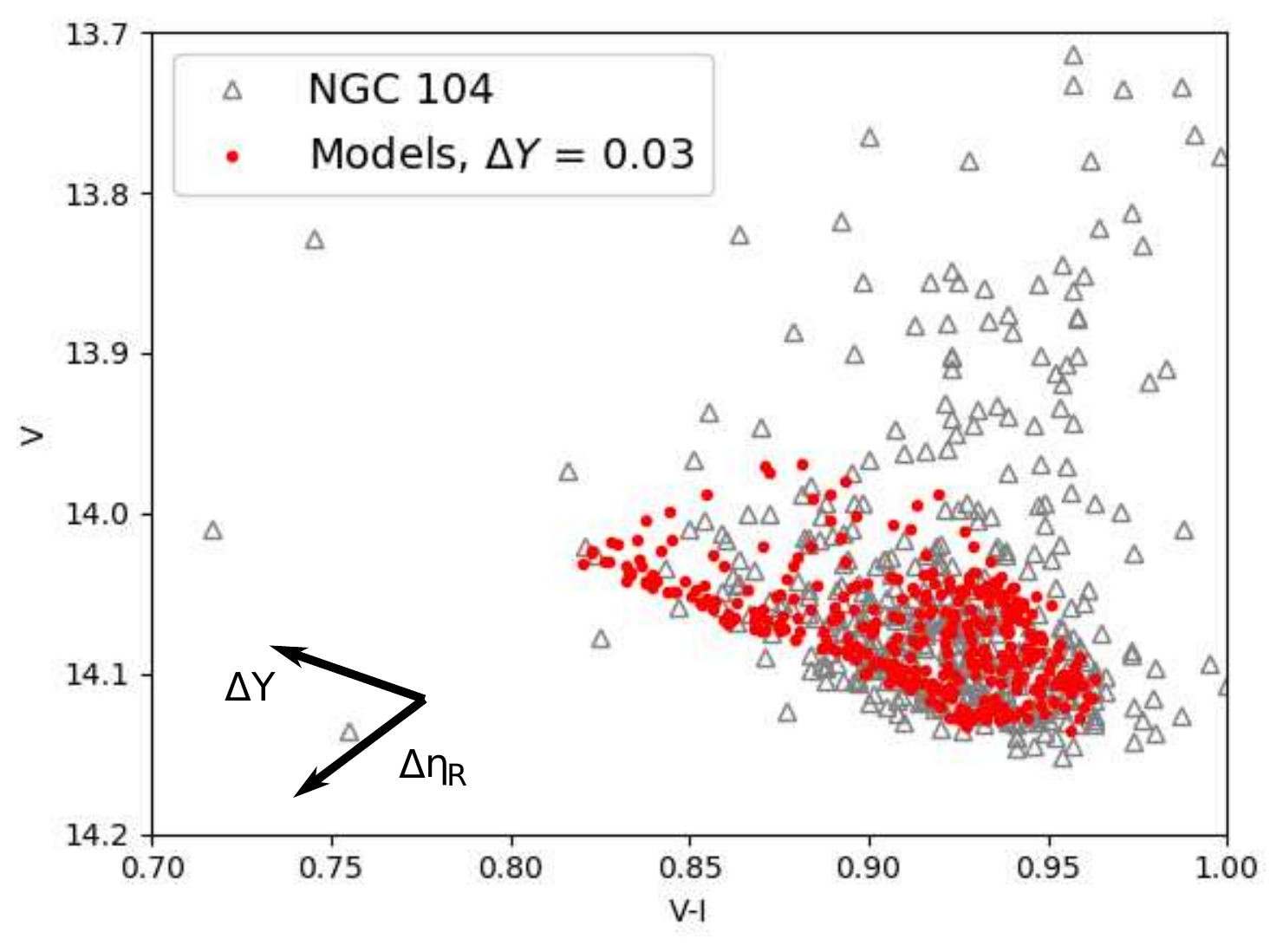} \\ 
    \caption{$VI$ CMD of NGC~104 HB (grey open triangles) together with our synthetic HB models (red circles). \textbf{\textit{Left:}} synthetic HB with $\Delta Y$=0. 
\textbf{\textit{Right:}} Synthetic HB with  $\Delta Y$=0.03 and uniform helium abundance distribution. The number of synthetic and observed stars in the box (blue) delimiting the HB region is the same. The arrows describe the direction along which variations in Y and mass-loss work, the amplitude being arbitrary here.} \label{Figure:NGC104}
\end{figure*}

We employed the same data \citep[$BVI$ photometry by][]{Bergbusch09} used by \cite{Salaris16}, an apparent distance modulus $(m-M)_V$ = 13.37 and reddening $E(B-V)$ = 0.04 \citep[][2010 edition]{Harris96}, 
and calculated $\alpha$-enhanced stellar models for [Fe/H] = -0.72, [$\alpha$/Fe]=+0.2, an age of 12~Gyr and various initial $Y$ and $\eta_R$. We use here for the extinction $A_B/A_V = 1.29719$ and $A_I/A_V= 0.60329$. Following the procedure described before, we find $\eta_R$ = 0.34 ($\Delta M_{RGB}\sim 0.17$~M$_\odot$) and $\Delta Y$ = 0.03 from the match of the observed HB. Figure~\ref{Figure:NGC104} compares the observed HB with synthetic HBs calculated with $\eta_R$ = 0.34 and both $\Delta Y$ = 0 (left-hand panel) and  $\Delta Y$ = 0.03 (right-hand panel).
A $\Delta Y$ of only 0.025 produces a HB too short and $\Delta Y$=0.035 produces a HB slightly too extended compared to the observations,

Our derived $\Delta Y$ = 0.03$\pm$0.005 is in good agreement with what is found in 
the literature.  

\section{Analysis of the Magellanic Clouds' cluster sample}\label{analysis}

\subsection{NGC~121}

\begin{figure*}
   \centering
  \includegraphics[width=0.35\textwidth]{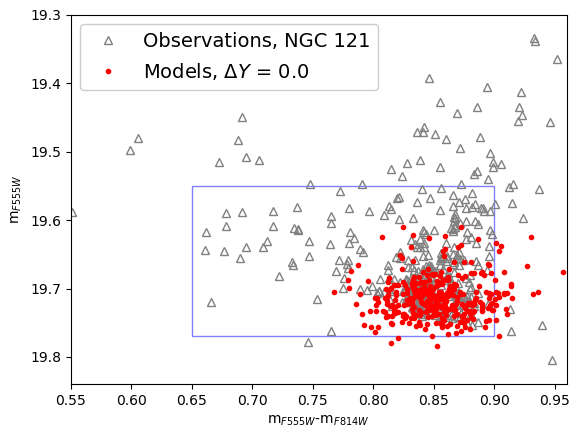}  
  \includegraphics[width=0.35\textwidth]{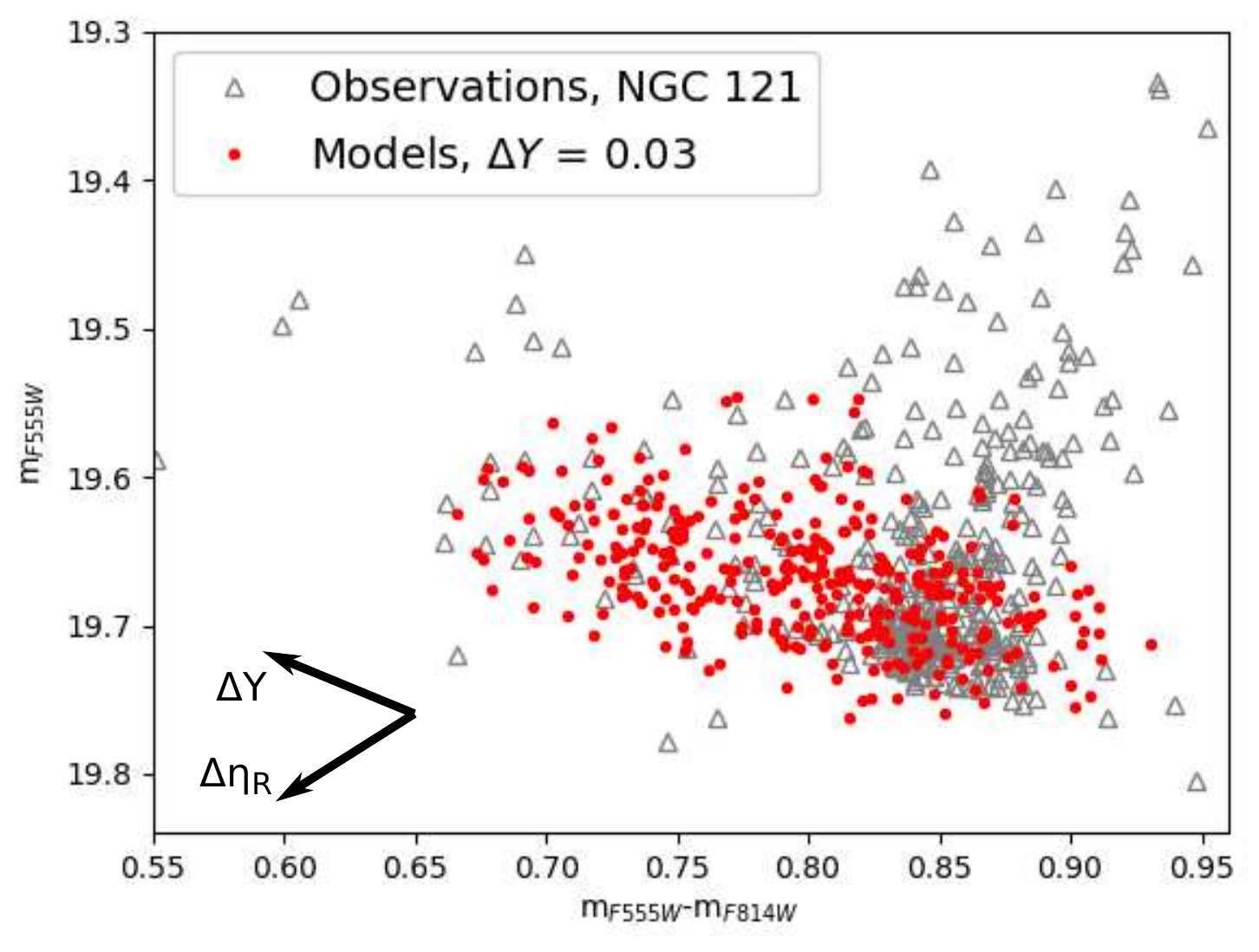}  \\    
  \includegraphics[width=0.35\textwidth]{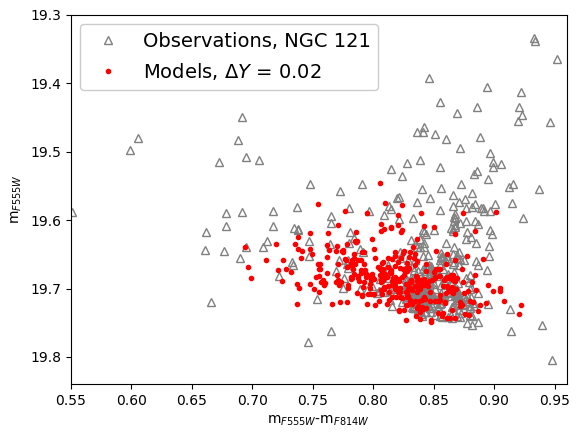}  
  \includegraphics[width=0.35\textwidth]{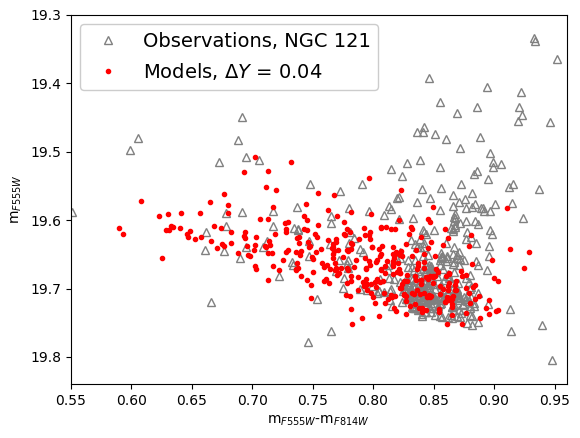} \\ 
    \caption{CMD of NGC~121 HB. Observations are represented by grey open triangles. Our synthetic HB models are overplotted in red. 
      \textbf{\textit{Top panels:}} Synthetic HB calculated with $\Delta Y$=0.0 and $\Delta Y$=0.03, respectively, both assuming $\eta_R$ = 0.33. The arrows describe the direction along which variations in Y and mass-loss work, the amplitude being arbitrary here.
    \textbf{\textit{Bottom panels:}} Synthetic HBs with $\eta_R$ = 0.33 calculated with $\Delta Y$=0.02 and $\Delta Y$=0.04 (see text for details).} \label{Figure:N121}
\end{figure*}

NGC~121 (SMC) has been investigated by \citet{Dalessandro16} and \citet{Niederhofer17_121}. The latter found two distinct populations from the analysis of the RGB with appropriate filter combinations, and 
they also concluded that a He abundance spread $\Delta Y$=0.025$\pm$0.005 is needed to explain the morphology of the cluster HB. 
Therefore this cluster, with properties very similar to massive Milky Way GCs (total mass $\sim 5.8\times10^{5} M_\odot$, age equal to $\sim$10.5~Gyr, and [Fe/H] = $-$1.30),   
allows us to compare again our results with previous independent results.  
We found that $\eta_R$ = 0.33 (corresponding to a total RGB mass loss $\Delta M_{RGB}\sim 0.145$~M$_\odot$, irrespective of the initial $Y$ of the models) 
and  $\Delta Y \sim$0.03 are required to match the colour extension and slope of the observed HB, as shown in Fig.~\ref{Figure:N121}.
A variation of $\eta_R$ at constant initial $Y$ would extend the synthetic HB orthogonally compared to the observations (see Fig.~\ref{Figure:Theory}). 
The derived $\Delta Y$ is consistent with \citet{Niederhofer17_121} result, based on a different set of HB stellar evolution models.

In the same Fig.~\ref{Figure:N121} we display the effect of changing $\Delta Y$ of the synthetic HBs by $\pm$0.01 around $\Delta Y =$0.03. 
It is obvious that in this case the colour extension of the observed HB is clearly not matched by the synthetic stars, implying that the error on our
estimates of $\Delta Y$ is lower than 0.01. This is the typical upper limit to the error in the $\Delta Y$ values obtained for the other clusters in our sample.

\begin{figure*}
   \centering
  \includegraphics[width=0.35\textwidth]{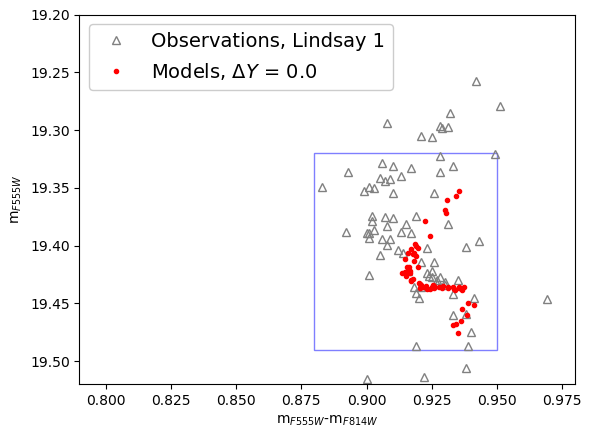}  
  \includegraphics[width=0.35\textwidth]{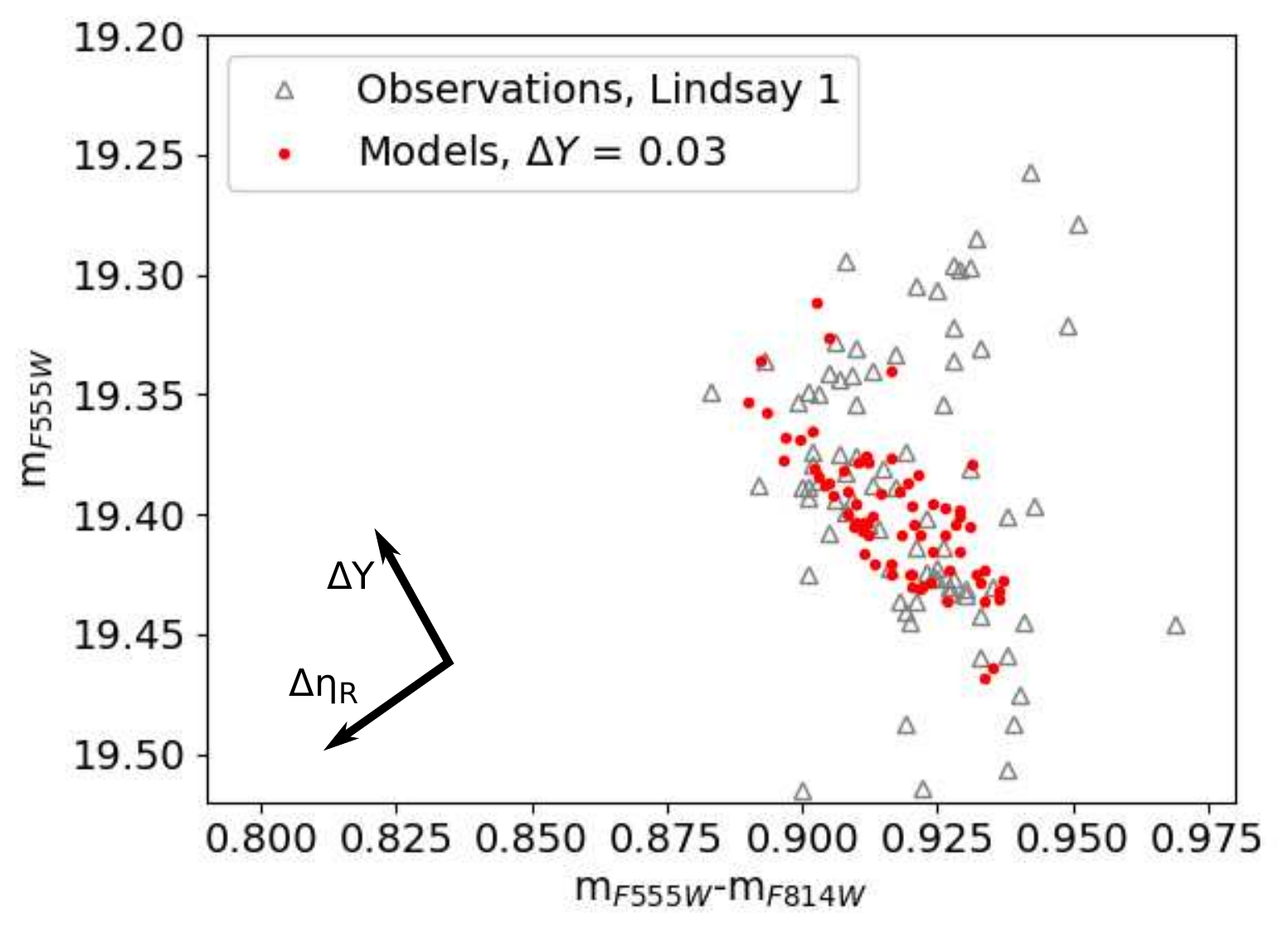} \\ 
  \includegraphics[width=0.35\textwidth]{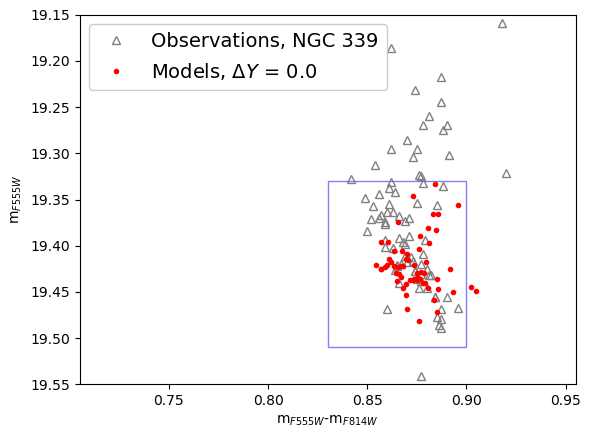}  
  \includegraphics[width=0.35\textwidth]{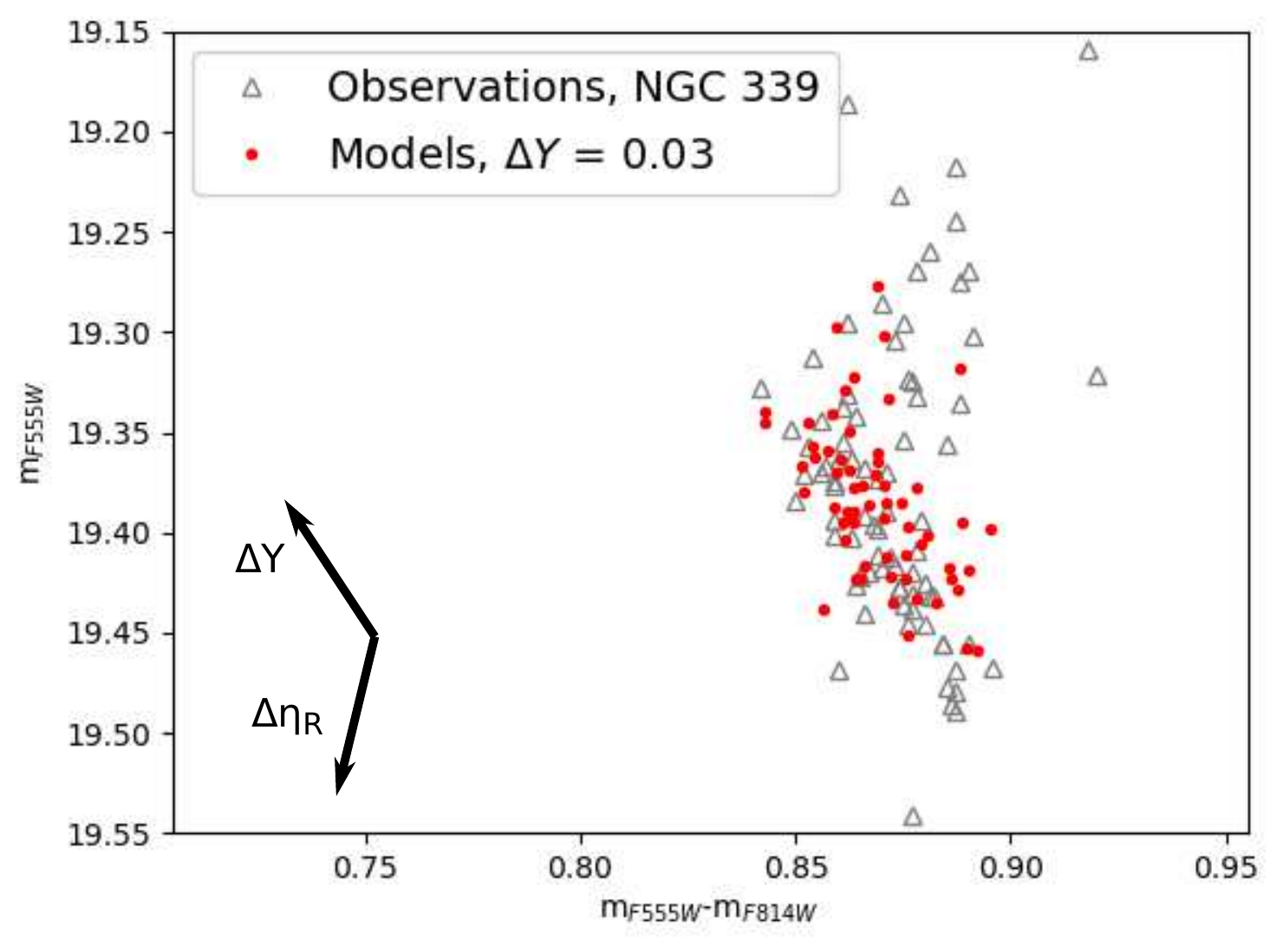} \\   
    \caption{CMDs of Lindsay~1, and NGC~339 HBs. Observations are represented by grey open triangles. Our synthetic HB models are overplotted in red. 
\textbf{\textit{From left to right:}} synthetic HB models at constant $Y$ and best fit $\eta_R$, and models with both best fit $\Delta Y$ and $\eta_R$, respectively. The arrows describe the direction along which variations in Y and mass-loss work, the amplitude being arbitrary here.} \label{Figure:AllY}
\end{figure*}

\begin{figure*}
   \centering
  \includegraphics[width=0.35\textwidth]{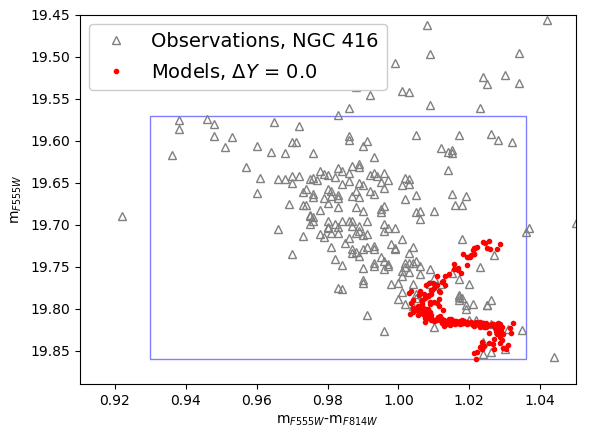}  
  \includegraphics[width=0.35\textwidth]{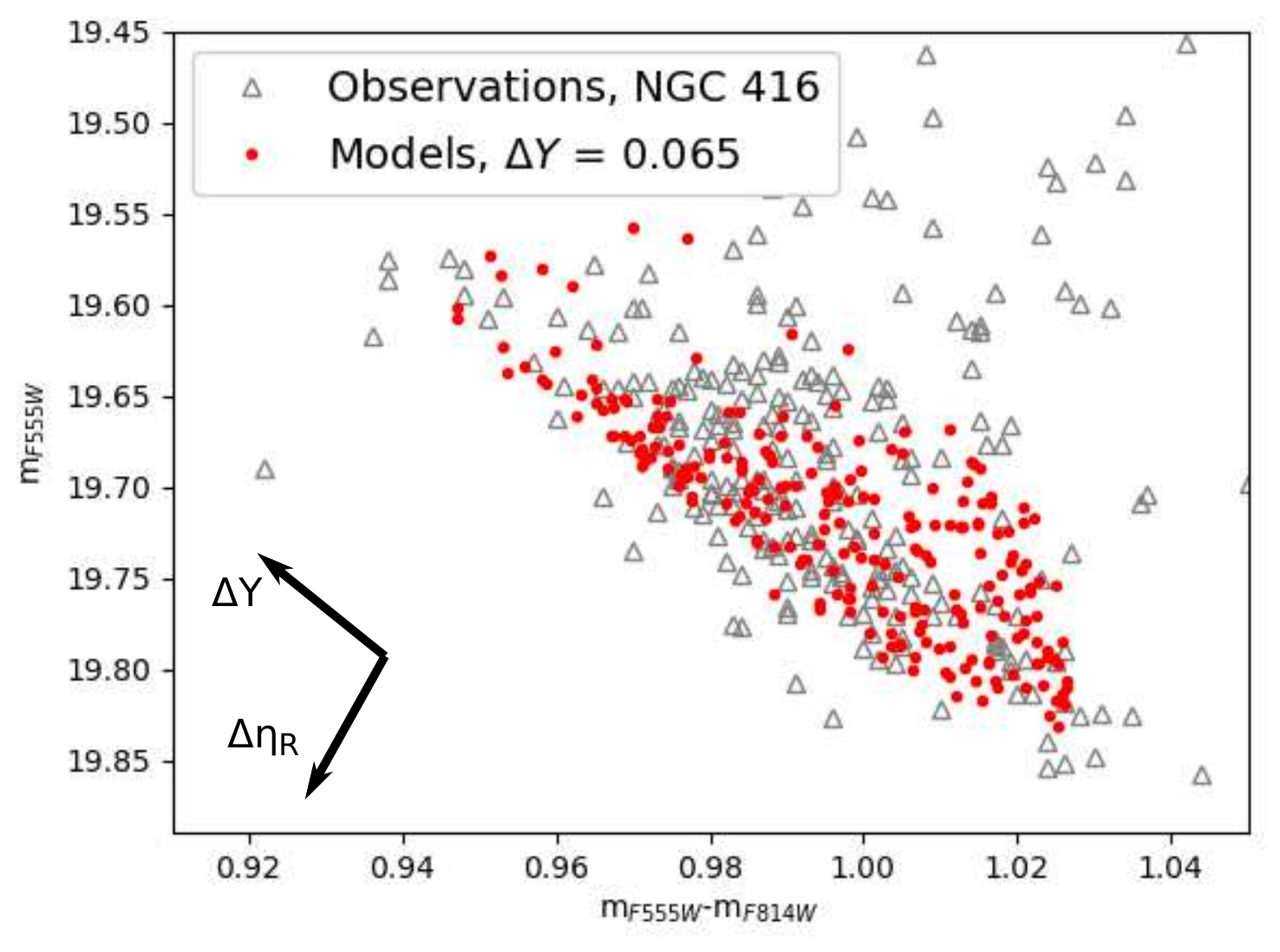} \\    
  \includegraphics[width=0.35\textwidth]{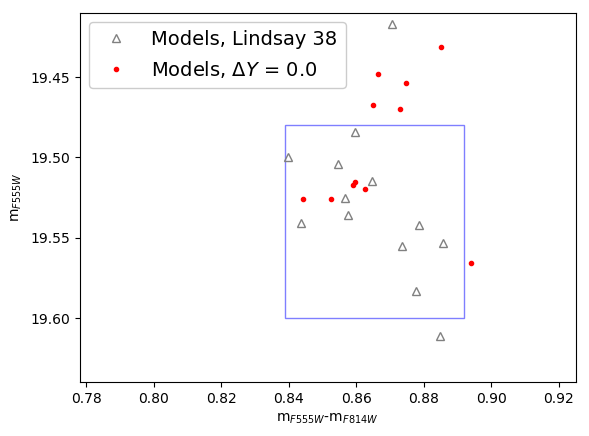}  
  \includegraphics[width=0.35\textwidth]{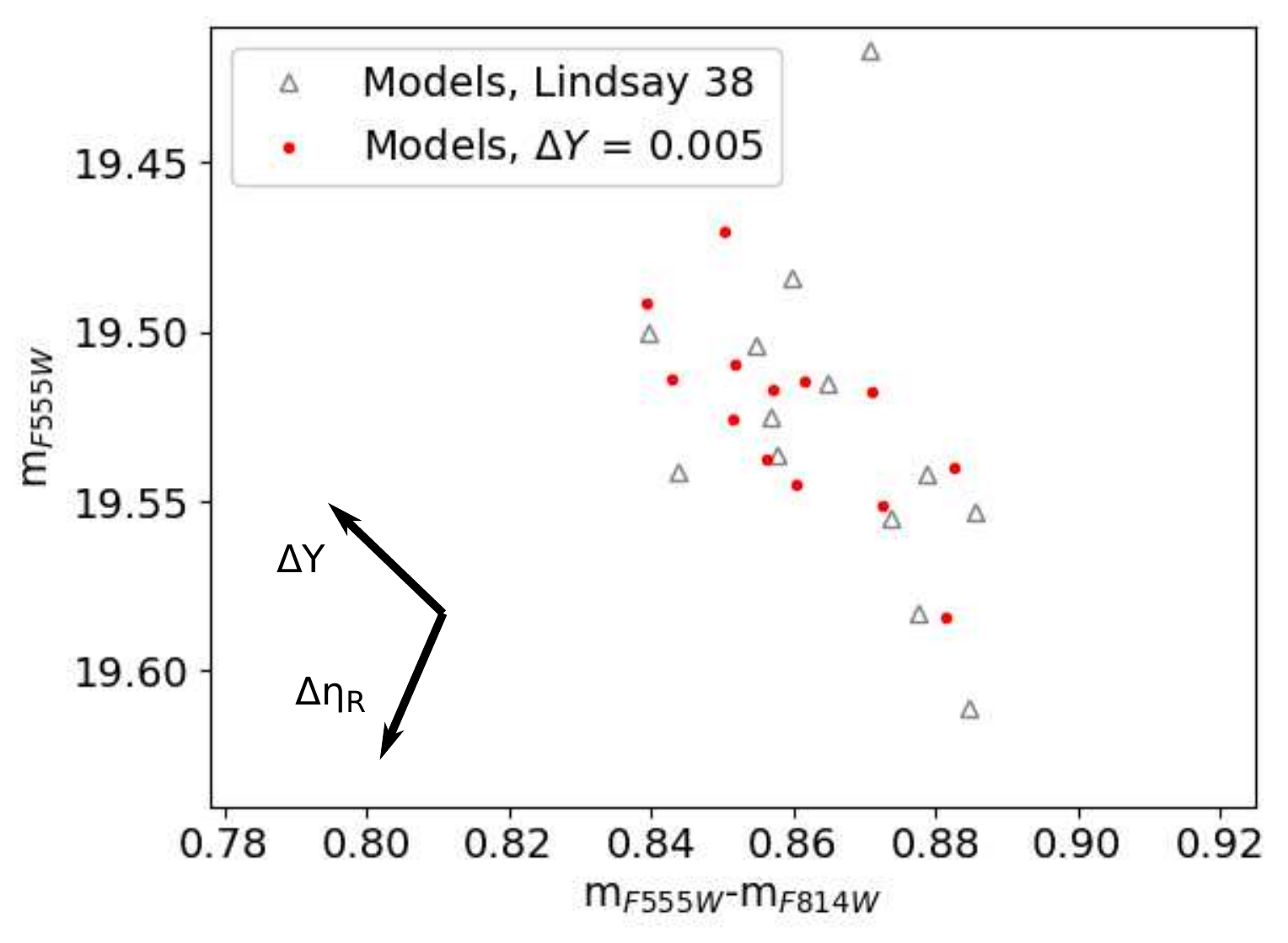} \\ 
  \includegraphics[width=0.35\textwidth]{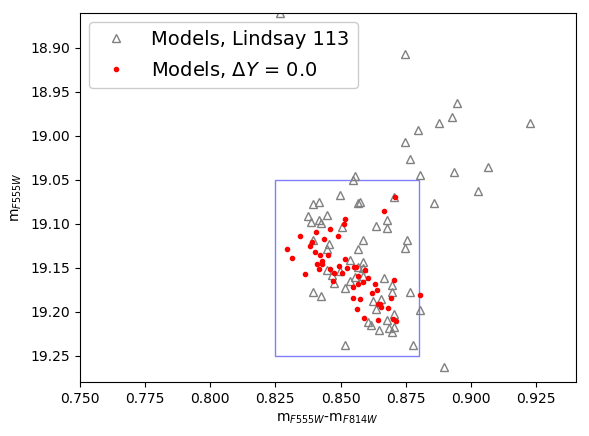}  
  \includegraphics[width=0.35\textwidth]{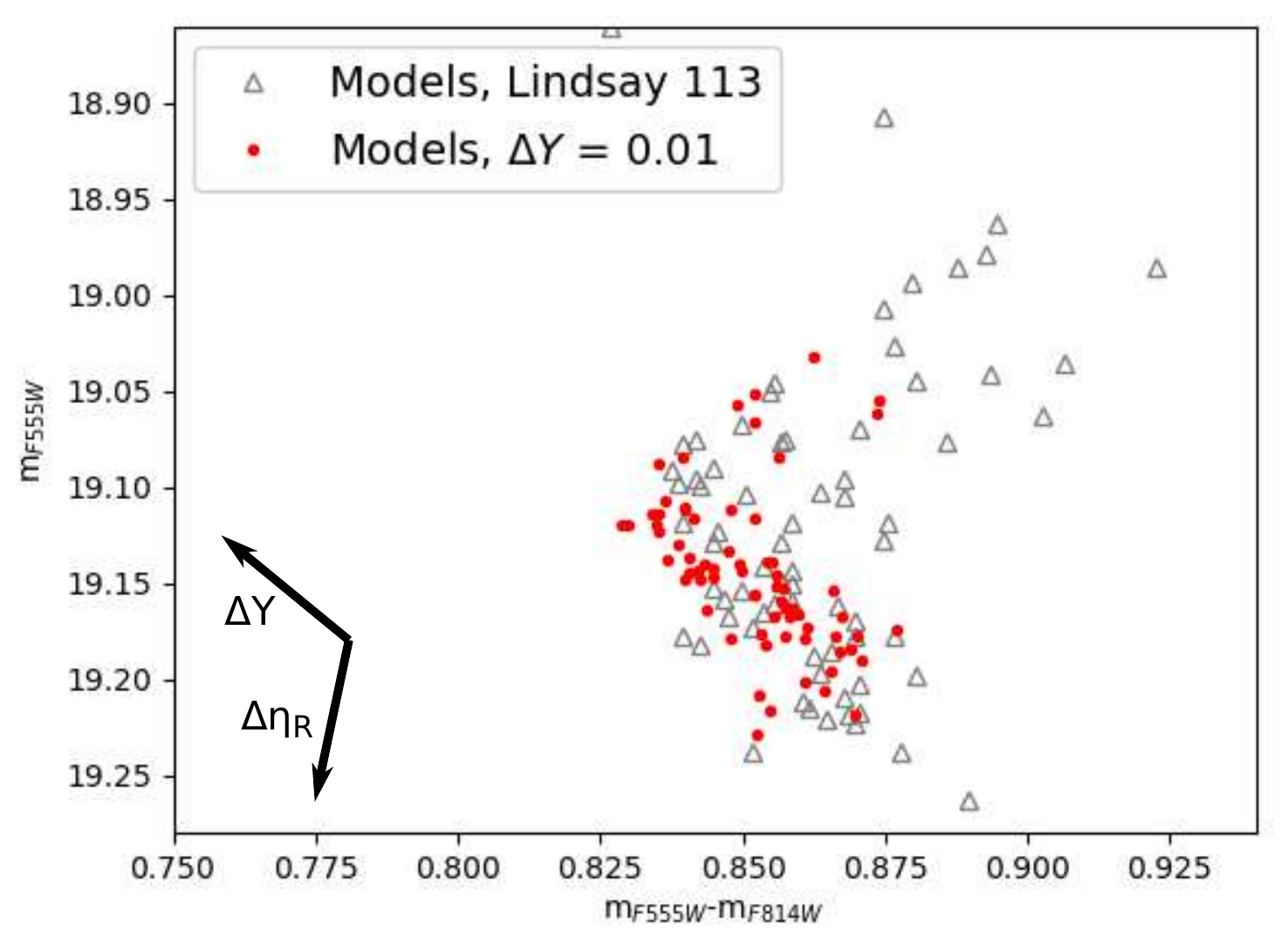} \\    
    \caption{As Fig.~\ref{Figure:AllY} but for NGC~416, Lindsay~38 and Lindsay~113. The arrows describe the direction along which variations in Y and mass-loss work, the amplitude being arbitrary here.} \label{Figure:AllY2}
\end{figure*}

\subsection{Lindsay~1}

Lindsay~1 (SMC), has a mass and metallicity typical of a Galactic  GC ($1.7\times10^{5} M_\odot$, [Fe/H] =$-$1.14 respectively) but a younger age ($\sim$7.5~Gyr).
\cite{Hollyhead17} found a significant nitrogen abundance spread ($\Delta$[N/Fe]$>$ 1~dex) among stars located below the RGB bump, a signature of
GC-like multiple stellar populations.
Later, \cite{Niederhofer17} detected a photometric split of the RGB in suitable photometric filter combinations, a signature of a N spread among its stars.

We determine from our HB fitting procedure $\eta_R$ = 0.3 ($\Delta M_{RGB}\sim 0.11$~M$_\odot$) and $\Delta Y\sim$0.03 (see Fig.~\ref{Figure:AllY}).

\subsection{NGC~339}

NGC~339 is a SMC cluster with total mass equal to 2.9$\times10^{5}$~M$_\odot$, an age of $\sim$6~Gyr, and [Fe/H] =$-$1.12.
\cite{Niederhofer17} found a photometric RGB splitting, characteristic of the presence of the multiple stellar populations.
From the HB fitting we determine $\eta_R$ = 0.4 ($\Delta M_{RGB}\sim 0.14$~M$_\odot$) and $\Delta Y \sim 0.03$ (see Fig.~\ref{Figure:AllY}).

\subsection{NGC~416}

NGC~416 is a SMC cluster very similar to NGC~339, with a total mass equal to $2.3\times10^{5} M_\odot$, an age $\sim$6~Gyr, and [Fe/H] =$-$1.00. 
We use here the data from \cite{Niederhofer17} corrected for differential reddening, that affects this cluster.
\citet{Niederhofer17} found also in this cluster a RGB splitting, signature of the presence of multiple stellar populations.

Our HB fitting provides $\eta_R$ = 0.4 ($\Delta M_{RGB}\sim 0.145$~M$_\odot$) and $\Delta Y$ = 0.065 (see Fig.~\ref{Figure:AllY2}).
This range of initial $Y$ is much larger than in the previous clusters, and might be at least 
slightly overestimated if there is some residual differential reddening not accounted for, given that the reddening vector is aligned with the HB slope.

\subsection{Lindsay 38}

The SMC cluster Lindsay~38 has an age similar to NGC~416 and NGC~339 ($\sim$6~Gyr), a lower mass ($\sim 3.35\times10^{4} M_\odot$) and a lower metallicity ([Fe/H] = $-$1.50).
The HB fitting provides $\eta_R$ = 0.3  ($\Delta M_{RGB}\sim 0.09$~M$_\odot$), but there is no strong indication of $\Delta Y>$0.
Fig.~\ref{Figure:AllY2} shows that $\Delta Y$=0.005 is probably an upper limit to the range of initial He in this cluster.

\subsection{Lindsay 113}

Lindsay~113 is the youngest SMC cluster in our sample ($\sim$4.5~Gyr), the least massive one ($\sim 2.3\times10^{4} M_\odot$), and metal-poor ([Fe/H] = $-$1.40).
We derive from the HB fitting $\eta_R$ = 0.3 ($\Delta M_{RGB}\sim 0.08$~M$_\odot$), and again no strong signature of a helium abundance spread.
Figure~\ref{Figure:AllY2} shows that $\Delta Y\sim$0.01 is very likely an upper limit to the possible $Y$ spread amongst the cluster stars.

\begin{figure*}
   \centering
  \includegraphics[width=0.3\textwidth]{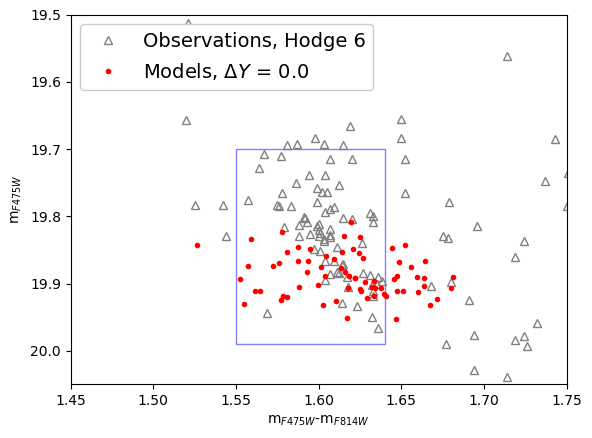}
  \includegraphics[width=0.3\textwidth]{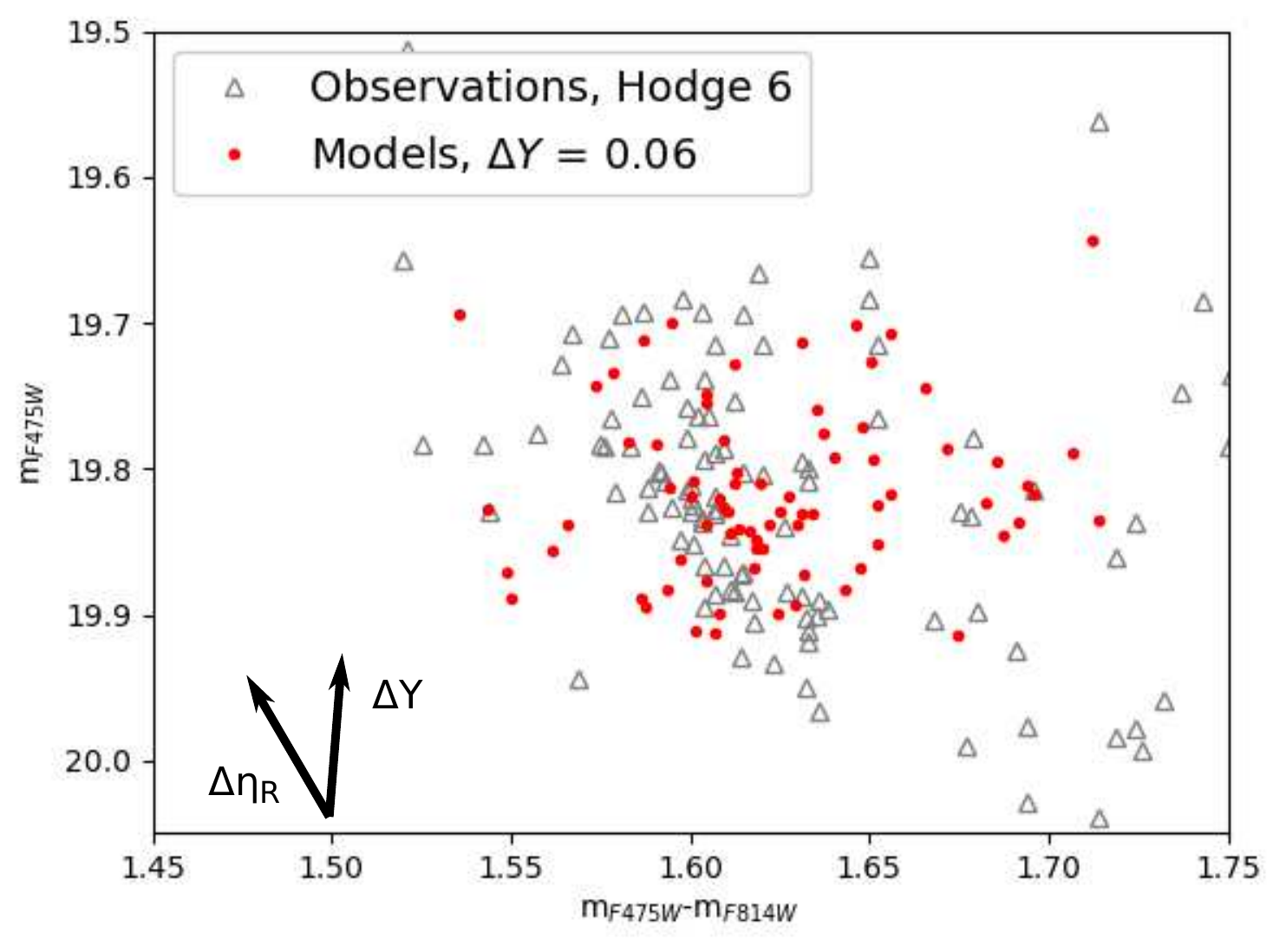}   
  \includegraphics[width=0.3\textwidth]{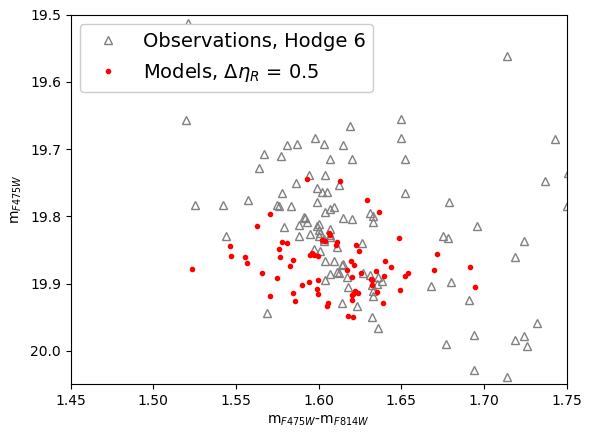} \\ 
  \includegraphics[width=0.3\textwidth]{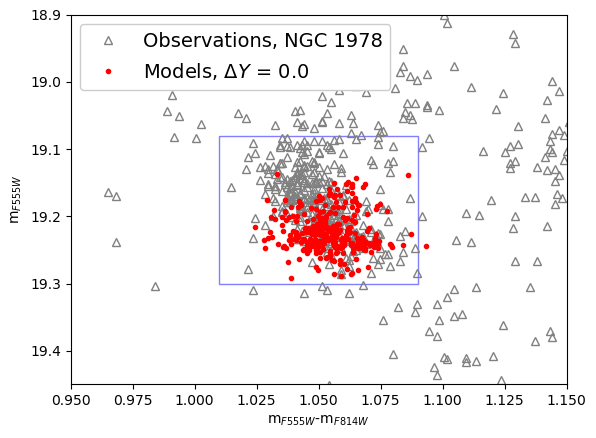}
  \includegraphics[width=0.3\textwidth]{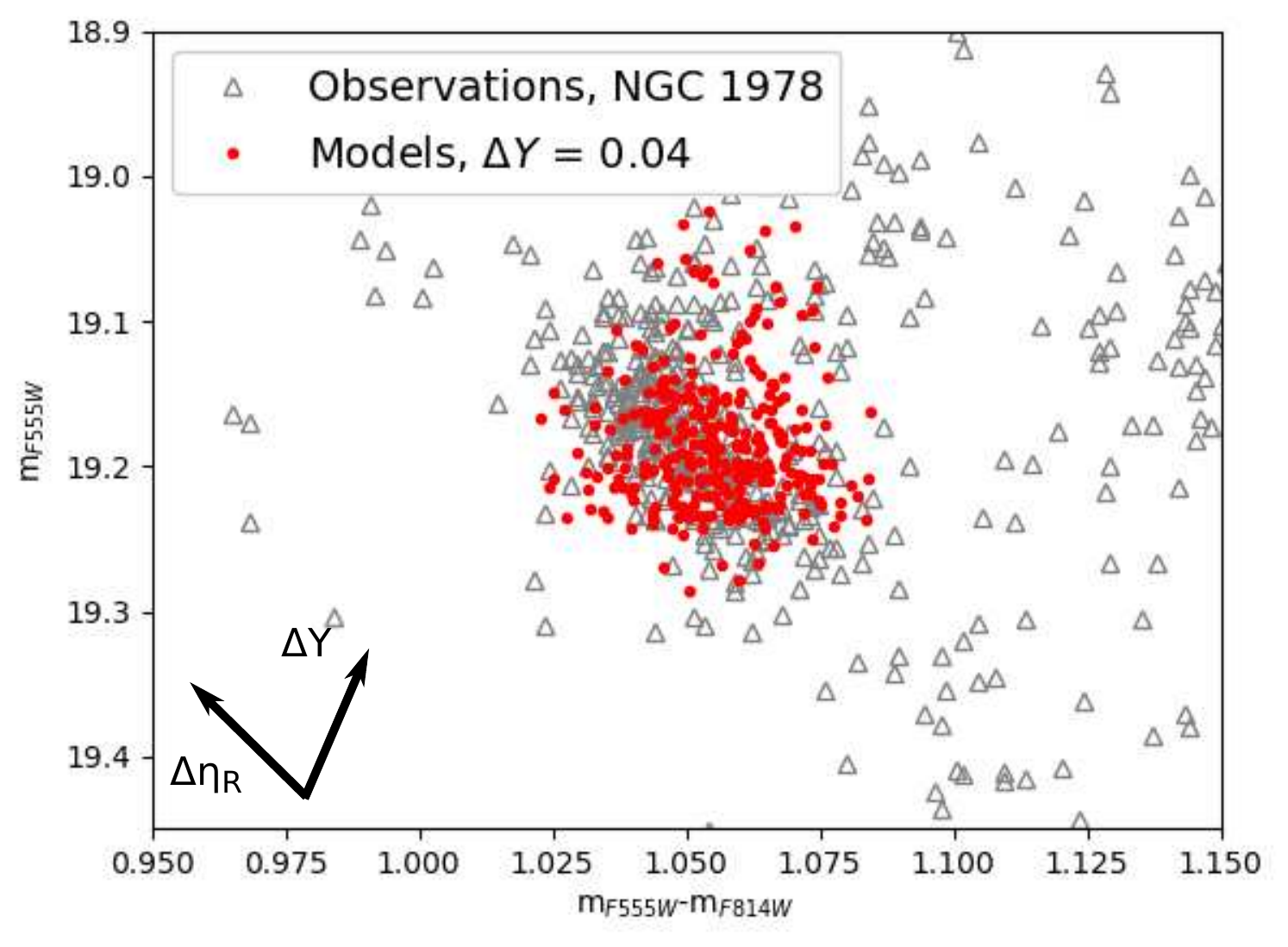}
  \includegraphics[width=0.3\textwidth]{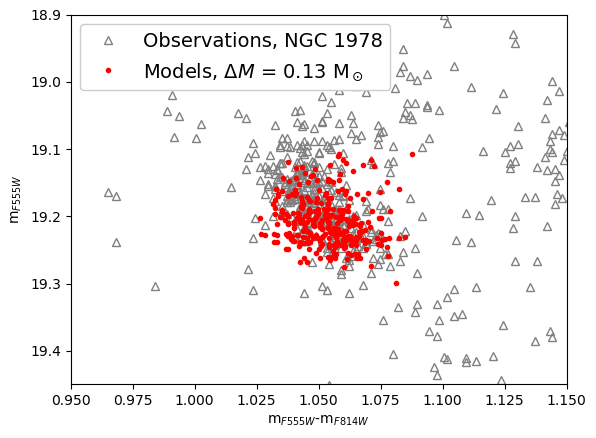} \\   
  \caption{CMDs of Hodge~6 and NGC~1978, the two youngest clusters in our sample. Observations are denoted with grey open triangles, synthetic HB
    models are overplotted in red. \textbf{\textit{From left to right:}} $\Delta Y$=0 models, models with $\Delta Y>$0 ($\eta_R$ fixed to the best fit value used in
    the left panel), and with $\Delta \eta_R >0$ ($Y$ fixed to the He-normal value of the left panel). The arrows describe the direction along which variations in Y and mass-loss work, the amplitude being arbitrary here.} \label{Figure:young}
\end{figure*}

\subsection{Hodge~6 and NGC~1978}\label{1978}

These two LMC clusters are the youngest clusters in our sample, with ages equal to $\sim$2.25 (Hodge~6) and $\sim$1.9~Gyr (NGC~1978), and [Fe/H] around $-$0.40
(see Table~\ref{Table:ClustersHB}). Multiple populations have been found in both clusters \citep[][Hollyhead et al. \textit{submitted}]{Martocchia18_1978}.

Due to their younger age, the direction of the $\Delta \eta_R$ and $\Delta Y$ vectors
is different compared to the case of the other clusters, as shown in Fig.~\ref{Figure:Theory}. 
The different direction of these two vectors compared to the older clusters, coupled to the photometric error of these observations --
of the order of 0.01-0.03~mag in magnitudes and
colours-- makes it difficult to reach a definitive conclusion about the existence  of a $\Delta Y>$0 in these two clusters.
Figure.~\ref{Figure:young} shows that an initial He spread (at fixed mass loss) or a mass loss spread (at fixed $Y$)
can similarly approximate the colour extension and slope of the observed CMD of core He burning stars.

If we make the assumption that $\eta_R$ must be constant, in agreement with the results for the other clusters in our sample, we would obtain
$\Delta Y\sim$0.06 for Hodge~6, and $\Delta Y\sim$0.04 for NGC~1978. But without this assumption, the CMD analysis does not discriminate between
a spread in $\eta_R$ or in $Y$ for these two clusters. However, we also note that Hodge~6 has the largest photometric errors of any of the clusters in our sample,  adding further uncertainty for this cluster.

\section{Discussion}\label{discussion}

We have determined the total initial He abundance spread $\Delta Y$ in a sample of intermediate-age, massive LMC and SMC clusters -- and the old cluster NGC~121--
by reproducing the shape and colour extension of their HB/RC stars with synthetic HB models.
Our derived $\Delta Y$ values are shown in Table~\ref{Table:results}. The typical error on these estimates of $\Delta Y$ is below 0.01.

\begin{figure}
   \centering
   \includegraphics[width=0.49\textwidth]{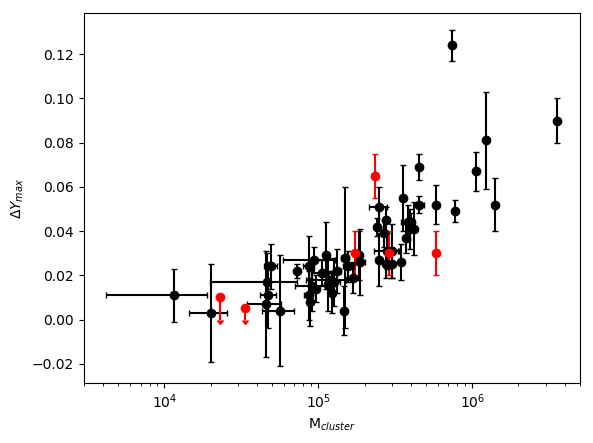}   
   \caption{Relation between $\Delta Y$ and cluster mass (in solar mass units). Galactic GC data from \protect\cite{Milone18} and \protect\cite{Baumgardt18} are displayed in black circles, the results for our SMC and LMC clusters are displayed as red circles.}
   \label{Figure:discussion}
\end{figure}

We can compare our results with the estimates by \citet{Lagioia19}. These authors found spreads of initial He abundances equal to 0.009$\pm$0.006, 0.007$\pm$0.004, 0.010$\pm$0.003, 0.000$\pm$0.004
for NGC~121, NGC~339, NGC~416 and Lindsay~1, respectively. These values are clearly smaller than our results in Table~\ref{Table:results}. But as mentioned already in the Introduction,
the method employed by \citet{Lagioia19} most likely determines mean abundance spreads among the cluster subpopulations, whereas our modelling tends to determine the maximum
abundance spread, irrespective of the exact distribution of initial He abundances. This is quite clear by looking at the HB of NGC~121 in Fig.~\ref{Figure:N121}. The bulk of the HB
population has ($m_{\mathrm F555W}-m_{\mathrm F814W}$)$>$0.8, consistent with a negligible $\Delta Y$ with just a plume of stars extending towards bluer colors and brighter magnitudes, that is, with
significantly different initial $Y$.

The values in Table~\ref{Table:results} are also plotted in Fig.~\ref{Figure:discussion} as a function of the mass of the host cluster. In the same figure we display also
the maximum initial $Y$ spread determined for a sample of Galactic GCs by \citet{Milone18}.
\citet{Milone18} found a trend between $\Delta Y$ and the mass of the host cluster, that is clearly visible in Fig.~\ref{Figure:discussion} and the results for our clusters follow this trend well. We found a Spearman rank-order correlation coefficient of 0.64 (p-value $\sim$ 0.17) between $\Delta Y_{max}$ and the logarithm of the cluster mass. This result confirms the ubiquity of multiple stellar populations in massive
intermediate-age clusters and GCs, questioning at the same time the distinction between these two classes of stellar systems.
Interestingly, our very tentative determination of $\Delta Y$ for NGC~1978 would fit the trend of Galactic GCs, whereas the $\Delta Y$ for Hodge~6 would be much higher for its value of total mass (but note that this final measurement is highly uncertain due to the photometric errors and age of the cluster as discussed in Sec.~\ref{1978}).

We also searched for possible trends of $\Delta Y$ with the cluster age amongst our cluster sample, but we did not find any statistically significant correlation
(Spearman rank-order correlation coefficient of 0.35, p-value $\sim$ 0.49, between $\Delta Y_{max}$ and M$_{cluster}$).
This result is, to some degree, surprising given that the N abundance spreads has been found correlated with age in MCs intermediate-age clusters \citep{Martocchia18,Martocchia18_1978},
and may potentially shine a new light on the MP phenomenon.

\begin{table}
\centering
\begin{tabular}{| c | c | c | c |}
	\hline \hline
    %& \multicolumn{4} {c} {Input} \\
    ID & $\Delta Y_{max}$ \\ \hline \hline
    NGC~121 & 0.03  \\
    Lindsay~1 & 0.03  \\
    NGC~339 & 0.03  \\
    NGC~416 & 0.065 \\     
    Lindsay~38 & $\le$0.005  \\       
    Lindsay~113 & $\le$0.01 \\    
    Hodge 6 &  (0.06) \\
    NGC~1978 &  (0.04) \\ 
    \hline \hline
\end{tabular}
\caption[]{Initial helium abundance spread derived from the HB fitting. The values in parenthesis are determined in the assumption that the RGB mass loss does
not vary among clusters' stars, because of a degeneracy between thr effects of mass loss spread and He spread in these clusters (see text for details).} \label{Table:results}
\end{table}

\section*{Acknowledgements}
We warmly thank E. Dalessandro for useful discussions. W. Chantereau acknowledges funding from the Swiss National Science Foundation under grant P2GEP2\_171971. 
N.B. and W.C. gratefully acknowledge financial support from the European Research Council (ERC-CoG-646928, Multi-Pop). 
N.B. gratefully acknowledges financial support from the Royal Society (University Research Fellowship). Finally, we warmly thank the referee for the pertinent suggestions that have helped us improve the presentation of our results.

\bibliographystyle{mnras}
\bibliography{main}

\begin{thebibliography}{}
\makeatletter
\relax
\def\mn@urlcharsother{\let\do\@makeother \do\$\do\&\do\#\do\^\do\_\do\%\do\~}
\def\mn@doi{\begingroup\mn@urlcharsother \@ifnextchar [ {\mn@doi@}
  {\mn@doi@[]}}
\def\mn@doi@[#1]#2{\def\@tempa{#1}\ifx\@tempa\@empty \href
  {http://dx.doi.org/#2} {doi:#2}\else \href {http://dx.doi.org/#2} {#1}\fi
  \endgroup}
\def\mn@eprint#1#2{\mn@eprint@#1:#2::\@nil}
\def\mn@eprint@arXiv#1{\href {http://arxiv.org/abs/#1} {{\tt arXiv:#1}}}
\def\mn@eprint@dblp#1{\href {http://dblp.uni-trier.de/rec/bibtex/#1.xml}
  {dblp:#1}}
\def\mn@eprint@#1:#2:#3:#4\@nil{\def\@tempa {#1}\def\@tempb {#2}\def\@tempc
  {#3}\ifx \@tempc \@empty \let \@tempc \@tempb \let \@tempb \@tempa \fi \ifx
  \@tempb \@empty \def\@tempb {arXiv}\fi \@ifundefined
  {mn@eprint@\@tempb}{\@tempb:\@tempc}{\expandafter \expandafter \csname
  mn@eprint@\@tempb\endcsname \expandafter{\@tempc}}}

\bibitem[\protect\citeauthoryear{{Anderson}, {Piotto}, {King}, {Bedin}  \&
  {Guhathakurta}}{{Anderson} et~al.}{2009}]{Anderson09}
{Anderson} J.,  {Piotto} G.,  {King} I.~R.,  {Bedin} L.~R.,   {Guhathakurta}
  P.,  2009, \mn@doi [\apjl] {10.1088/0004-637X/697/1/L58}, \href
  {http://adsabs.harvard.edu/abs/2009ApJ...697L..58A} {697, L58}

\bibitem[\protect\citeauthoryear{{Asplund}, {Grevesse}, {Sauval}  \&
  {Scott}}{{Asplund} et~al.}{2009}]{Asplund09}
{Asplund} M.,  {Grevesse} N.,  {Sauval} A.~J.,   {Scott} P.,  2009, \mn@doi
  [\araa] {10.1146/annurev.astro.46.060407.145222}, \href
  {http://adsabs.harvard.edu/abs/2009ARA%26A..47..481A} {47, 481}

\bibitem[\protect\citeauthoryear{{Bastian} \& {Lardo}}{{Bastian} \&
  {Lardo}}{2018}]{Bastian18}
{Bastian} N.,  {Lardo} C.,  2018, \mn@doi [\araa]
  {10.1146/annurev-astro-081817-051839}, \href
  {http://adsabs.harvard.edu/abs/2018ARA%26A..56...83B} {56, 83}

\bibitem[\protect\citeauthoryear{{Baumgardt} \& {Hilker}}{{Baumgardt} \&
  {Hilker}}{2018}]{Baumgardt18}
{Baumgardt} H.,  {Hilker} M.,  2018, \mn@doi [\mnras] {10.1093/mnras/sty1057},
  \href {http://adsabs.harvard.edu/abs/2018MNRAS.478.1520B} {478, 1520}

\bibitem[\protect\citeauthoryear{{Bergbusch} \& {Stetson}}{{Bergbusch} \&
  {Stetson}}{2009}]{Bergbusch09}
{Bergbusch} P.~A.,  {Stetson} P.~B.,  2009, \mn@doi [\aj]
  {10.1088/0004-6256/138/5/1455}, \href
  {http://adsabs.harvard.edu/abs/2009AJ....138.1455B} {138, 1455}

\bibitem[\protect\citeauthoryear{{Busso} et~al.,}{{Busso}
  et~al.}{2007}]{Busso07}
{Busso} G.,  et~al., 2007, \mn@doi [\aap] {10.1051/0004-6361:20077806}, \href
  {http://adsabs.harvard.edu/abs/2007A%26A...474..105B} {474, 105}

\bibitem[\protect\citeauthoryear{{Caloi} \& {D'Antona}}{{Caloi} \&
  {D'Antona}}{2007}]{Caloi07}
{Caloi} V.,  {D'Antona} F.,  2007, \mn@doi [\aap] {10.1051/0004-6361:20066074},
  \href {http://adsabs.harvard.edu/abs/2007A%26A...463..949C} {463, 949}

\bibitem[\protect\citeauthoryear{{Carretta}, {Bragaglia}, {Gratton},
  {Recio-Blanco}, {Lucatello}, {D'Orazi}  \& {Cassisi}}{{Carretta}
  et~al.}{2010}]{Carretta10}
{Carretta} E.,  {Bragaglia} A.,  {Gratton} R.~G.,  {Recio-Blanco} A.,
  {Lucatello} S.,  {D'Orazi} V.,   {Cassisi} S.,  2010, \mn@doi [\aap]
  {10.1051/0004-6361/200913451}, \href
  {http://adsabs.harvard.edu/abs/2010A%26A...516A..55C} {516, A55}

\bibitem[\protect\citeauthoryear{{Choi}, {Dotter}, {Conroy}, {Cantiello},
  {Paxton}  \& {Johnson}}{{Choi} et~al.}{2016}]{Choi16}
{Choi} J.,  {Dotter} A.,  {Conroy} C.,  {Cantiello} M.,  {Paxton} B.,
  {Johnson} B.~D.,  2016, \mn@doi [\apj] {10.3847/0004-637X/823/2/102}, \href
  {http://adsabs.harvard.edu/abs/2016ApJ...823..102C} {823, 102}

\bibitem[\protect\citeauthoryear{{Coc}, {Vangioni-Flam}, {Descouvemont},
  {Adahchour}  \& {Angulo}}{{Coc} et~al.}{2004}]{Coc04}
{Coc} A.,  {Vangioni-Flam} E.,  {Descouvemont} P.,  {Adahchour} A.,   {Angulo}
  C.,  2004, \mn@doi [\apj] {10.1086/380121}, \href
  {http://adsabs.harvard.edu/abs/2004ApJ...600..544C} {600, 544}

\bibitem[\protect\citeauthoryear{{Dalessandro}, {Salaris}, {Ferraro},
  {Cassisi}, {Lanzoni}, {Rood}, {Fusi Pecci}  \& {Sabbi}}{{Dalessandro}
  et~al.}{2011}]{Dalessandro11}
{Dalessandro} E.,  {Salaris} M.,  {Ferraro} F.~R.,  {Cassisi} S.,  {Lanzoni}
  B.,  {Rood} R.~T.,  {Fusi Pecci} F.,   {Sabbi} E.,  2011, \mn@doi [\mnras]
  {10.1111/j.1365-2966.2010.17479.x}, \href
  {http://adsabs.harvard.edu/abs/2011MNRAS.410..694D} {410, 694}

\bibitem[\protect\citeauthoryear{{Dalessandro}, {Salaris}, {Ferraro},
  {Mucciarelli}  \& {Cassisi}}{{Dalessandro} et~al.}{2013}]{Dalessandro13}
{Dalessandro} E.,  {Salaris} M.,  {Ferraro} F.~R.,  {Mucciarelli} A.,
  {Cassisi} S.,  2013, \mn@doi [\mnras] {10.1093/mnras/sts644}, \href
  {http://adsabs.harvard.edu/abs/2013MNRAS.430..459D} {430, 459}

\bibitem[\protect\citeauthoryear{{Dalessandro}, {Lapenna}, {Mucciarelli},
  {Origlia}, {Ferraro}  \& {Lanzoni}}{{Dalessandro}
  et~al.}{2016}]{Dalessandro16}
{Dalessandro} E.,  {Lapenna} E.,  {Mucciarelli} A.,  {Origlia} L.,  {Ferraro}
  F.~R.,   {Lanzoni} B.,  2016, \mn@doi [\apj] {10.3847/0004-637X/829/2/77},
  \href {http://adsabs.harvard.edu/abs/2016ApJ...829...77D} {829, 77}

\bibitem[\protect\citeauthoryear{{Di Criscienzo}, {Tailo}, {Milone},
  {D'Antona}, {Ventura}, {Dotter}  \& {Brocato}}{{Di Criscienzo}
  et~al.}{2015}]{diCriscienzo15}
{Di Criscienzo} M.,  {Tailo} M.,  {Milone} A.~P.,  {D'Antona} F.,  {Ventura}
  P.,  {Dotter} A.,   {Brocato} E.,  2015, \mn@doi [\mnras]
  {10.1093/mnras/stu2167}, \href
  {http://adsabs.harvard.edu/abs/2015MNRAS.446.1469D} {446, 1469}

\bibitem[\protect\citeauthoryear{{Glatt} et~al.,}{{Glatt}
  et~al.}{2008a}]{Glatt08_121}
{Glatt} K.,  et~al., 2008a, \mn@doi [\aj] {10.1088/0004-6256/135/4/1106}, \href
  {http://adsabs.harvard.edu/abs/2008AJ....135.1106G} {135, 1106}

\bibitem[\protect\citeauthoryear{{Glatt} et~al.,}{{Glatt}
  et~al.}{2008b}]{Glatt08}
{Glatt} K.,  et~al., 2008b, \mn@doi [\aj] {10.1088/0004-6256/136/4/1703}, \href
  {http://adsabs.harvard.edu/abs/2008AJ....136.1703G} {136, 1703}

\bibitem[\protect\citeauthoryear{{Glatt} et~al.,}{{Glatt}
  et~al.}{2011}]{Glatt11}
{Glatt} K.,  et~al., 2011, \mn@doi [\aj] {10.1088/0004-6256/142/2/36}, \href
  {http://adsabs.harvard.edu/abs/2011AJ....142...36G} {142, 36}

\bibitem[\protect\citeauthoryear{{Goudfrooij}, {Puzia}, {Kozhurina-Platais}  \&
  {Chandar}}{{Goudfrooij} et~al.}{2009}]{Goudfrooij09}
{Goudfrooij} P.,  {Puzia} T.~H.,  {Kozhurina-Platais} V.,   {Chandar} R.,
  2009, \mn@doi [\aj] {10.1088/0004-6256/137/6/4988}, \href
  {http://adsabs.harvard.edu/abs/2009AJ....137.4988G} {137, 4988}

\bibitem[\protect\citeauthoryear{{Goudfrooij} et~al.,}{{Goudfrooij}
  et~al.}{2014}]{Goudfrooij14}
{Goudfrooij} P.,  et~al., 2014, \mn@doi [\apj] {10.1088/0004-637X/797/1/35},
  \href {http://adsabs.harvard.edu/abs/2014ApJ...797...35G} {797, 35}

\bibitem[\protect\citeauthoryear{{Gratton} et~al.,}{{Gratton}
  et~al.}{2013}]{Gratton13}
{Gratton} R.~G.,  et~al., 2013, \mn@doi [\aap] {10.1051/0004-6361/201219976},
  \href {http://adsabs.harvard.edu/abs/2013A%26A...549A..41G} {549, A41}

\bibitem[\protect\citeauthoryear{{Harris}}{{Harris}}{1996}]{Harris96}
{Harris} W.~E.,  1996, \mn@doi [\aj] {10.1086/118116}, \href
  {http://adsabs.harvard.edu/abs/1996AJ....112.1487H} {112, 1487}

\bibitem[\protect\citeauthoryear{{Hollyhead} et~al.,}{{Hollyhead}
  et~al.}{2017}]{Hollyhead17}
{Hollyhead} K.,  et~al., 2017, \mn@doi [\mnras] {10.1093/mnrasl/slw179}, \href
  {http://adsabs.harvard.edu/abs/2017MNRAS.465L..39H} {465, L39}

\bibitem[\protect\citeauthoryear{{Hui-Bon-Hoa}, {LeBlanc}  \&
  {Hauschildt}}{{Hui-Bon-Hoa} et~al.}{2000}]{HuiBonHoa00}
{Hui-Bon-Hoa} A.,  {LeBlanc} F.,   {Hauschildt} P.~H.,  2000, \mn@doi [\apjl]
  {10.1086/312693}, \href {http://adsabs.harvard.edu/abs/2000ApJ...535L..43H}
  {535, L43}

\bibitem[\protect\citeauthoryear{{Lagarde}, {Decressin}, {Charbonnel},
  {Eggenberger}, {Ekstr{\"o}m}  \& {Palacios}}{{Lagarde}
  et~al.}{2012}]{Lagarde12}
{Lagarde} N.,  {Decressin} T.,  {Charbonnel} C.,  {Eggenberger} P.,
  {Ekstr{\"o}m} S.,   {Palacios} A.,  2012, \mn@doi [\aap]
  {10.1051/0004-6361/201118331}, \href
  {http://adsabs.harvard.edu/abs/2012A%26A...543A.108L} {543, A108}

\bibitem[\protect\citeauthoryear{{Lagioia}, {Milone}, {Marino}  \&
  {Dotter}}{{Lagioia} et~al.}{2019}]{Lagioia19}
{Lagioia} E.~P.,  {Milone} A.~P.,  {Marino} A.~F.,   {Dotter} A.,  2019,
  \mn@doi [\apj] {10.3847/1538-4357/aaf729}, \href
  {http://adsabs.harvard.edu/abs/2019ApJ...871..140L} {871, 140}

\bibitem[\protect\citeauthoryear{{Martocchia} et~al.,}{{Martocchia}
  et~al.}{2018a}]{Martocchia18}
{Martocchia} S.,  et~al., 2018a, \mn@doi [\mnras] {10.1093/mnras/stx2556},
  \href {http://adsabs.harvard.edu/abs/2018MNRAS.473.2688M} {473, 2688}

\bibitem[\protect\citeauthoryear{{Martocchia} et~al.,}{{Martocchia}
  et~al.}{2018b}]{Martocchia18_1978}
{Martocchia} S.,  et~al., 2018b, \mn@doi [\mnras] {10.1093/mnras/sty916}, \href
  {http://adsabs.harvard.edu/abs/2018MNRAS.477.4696M} {477, 4696}

\bibitem[\protect\citeauthoryear{{McDonald} \& {Zijlstra}}{{McDonald} \&
  {Zijlstra}}{2015}]{McDonald15}
{McDonald} I.,  {Zijlstra} A.~A.,  2015, \mn@doi [\mnras]
  {10.1093/mnras/stv007}, \href
  {http://adsabs.harvard.edu/abs/2015MNRAS.448..502M} {448, 502}

\bibitem[\protect\citeauthoryear{{Michaud}, {Richer}  \& {Richard}}{{Michaud}
  et~al.}{2011}]{Michaud11}
{Michaud} G.,  {Richer} J.,   {Richard} O.,  2011, \mn@doi [\aap]
  {10.1051/0004-6361/201015997}, \href
  {http://adsabs.harvard.edu/abs/2011A%26A...529A..60M} {529, A60}

\bibitem[\protect\citeauthoryear{{Milone} et~al.,}{{Milone}
  et~al.}{2012}]{Milone12}
{Milone} A.~P.,  et~al., 2012, \mn@doi [\apj] {10.1088/0004-637X/744/1/58},
  \href {http://adsabs.harvard.edu/abs/2012ApJ...744...58M} {744, 58}

\bibitem[\protect\citeauthoryear{{Milone} et~al.,}{{Milone}
  et~al.}{2018}]{Milone18}
{Milone} A.~P.,  et~al., 2018, \mn@doi [\mnras] {10.1093/mnras/sty2573}, \href
  {http://adsabs.harvard.edu/abs/2018MNRAS.481.5098M} {481, 5098}

\bibitem[\protect\citeauthoryear{{Mucciarelli}, {Ferraro}, {Origlia}  \& {Fusi
  Pecci}}{{Mucciarelli} et~al.}{2007}]{Mucciarelli07}
{Mucciarelli} A.,  {Ferraro} F.~R.,  {Origlia} L.,   {Fusi Pecci} F.,  2007,
  \mn@doi [\aj] {10.1086/513076}, \href
  {http://adsabs.harvard.edu/abs/2007AJ....133.2053M} {133, 2053}

\bibitem[\protect\citeauthoryear{{Niederhofer} et~al.,}{{Niederhofer}
  et~al.}{2017a}]{Niederhofer17_121}
{Niederhofer} F.,  et~al., 2017a, \mn@doi [\mnras] {10.1093/mnras/stw2269},
  \href {http://adsabs.harvard.edu/abs/2017MNRAS.464...94N} {464, 94}

\bibitem[\protect\citeauthoryear{{Niederhofer} et~al.,}{{Niederhofer}
  et~al.}{2017b}]{Niederhofer17}
{Niederhofer} F.,  et~al., 2017b, \mn@doi [\mnras] {10.1093/mnras/stw3084},
  \href {http://adsabs.harvard.edu/abs/2017MNRAS.465.4159N} {465, 4159}

\bibitem[\protect\citeauthoryear{{Piatti}, {Keller}, {Mackey}  \& {Da
  Costa}}{{Piatti} et~al.}{2014}]{Piatti14}
{Piatti} A.~E.,  {Keller} S.~C.,  {Mackey} A.~D.,   {Da Costa} G.~S.,  2014,
  \mn@doi [\mnras] {10.1093/mnras/stu1535}, \href
  {http://adsabs.harvard.edu/abs/2014MNRAS.444.1425P} {444, 1425}

\bibitem[\protect\citeauthoryear{{Reimers}}{{Reimers}}{1975}]{Reimers75}
{Reimers} D.,  1975, Memoires of the Societe Royale des Sciences de Liege,
  \href {http://adsabs.harvard.edu/abs/1975MSRSL...8..369R} {8, 369}

\bibitem[\protect\citeauthoryear{{Salaris}, {Weiss}, {Ferguson}  \&
  {Fusilier}}{{Salaris} et~al.}{2006}]{Salaris06}
{Salaris} M.,  {Weiss} A.,  {Ferguson} J.~W.,   {Fusilier} D.~J.,  2006,
  \mn@doi [\apj] {10.1086/504520}, \href
  {http://adsabs.harvard.edu/abs/2006ApJ...645.1131S} {645, 1131}

\bibitem[\protect\citeauthoryear{{Salaris}, {Cassisi}  \&
  {Pietrinferni}}{{Salaris} et~al.}{2016}]{Salaris16}
{Salaris} M.,  {Cassisi} S.,   {Pietrinferni} A.,  2016, \mn@doi [\aap]
  {10.1051/0004-6361/201628181}, \href
  {http://adsabs.harvard.edu/abs/2016A%26A...590A..64S} {590, A64}

\bibitem[\protect\citeauthoryear{{Sbordone}, {Salaris}, {Weiss}  \&
  {Cassisi}}{{Sbordone} et~al.}{2011}]{Sbordone11}
{Sbordone} L.,  {Salaris} M.,  {Weiss} A.,   {Cassisi} S.,  2011, \mn@doi
  [\aap] {10.1051/0004-6361/201116714}, \href
  {http://adsabs.harvard.edu/abs/2011A%26A...534A...9S} {534, A9}

\bibitem[\protect\citeauthoryear{{Westerlund}}{{Westerlund}}{1997}]{Westerlund97}
{Westerlund} B.~E.,  1997, {The Magellanic Clouds}

\bibitem[\protect\citeauthoryear{{Yong}, {Grundahl}  \& {Norris}}{{Yong}
  et~al.}{2015}]{Yong15}
{Yong} D.,  {Grundahl} F.,   {Norris} J.~E.,  2015, \mn@doi [\mnras]
  {10.1093/mnras/stu2334}, \href
  {http://adsabs.harvard.edu/abs/2015MNRAS.446.3319Y} {446, 3319}

\bibitem[\protect\citeauthoryear{{di Criscienzo}, {Ventura}, {D'Antona},
  {Milone}  \& {Piotto}}{{di Criscienzo} et~al.}{2010}]{diCriscienzo10}
{di Criscienzo} M.,  {Ventura} P.,  {D'Antona} F.,  {Milone} A.,   {Piotto} G.,
   2010, \mn@doi [\mnras] {10.1111/j.1365-2966.2010.17168.x}, \href
  {http://adsabs.harvard.edu/abs/2010MNRAS.408..999D} {408, 999}

\bibitem[\protect\citeauthoryear{{di Criscienzo} et~al.,}{{di Criscienzo}
  et~al.}{2011}]{diCriscienzo11}
{di Criscienzo} M.,  et~al., 2011, \mn@doi [\mnras]
  {10.1111/j.1365-2966.2011.18642.x}, \href
  {http://adsabs.harvard.edu/abs/2011MNRAS.414.3381D} {414, 3381}

\makeatother
\end{thebibliography}

\end{document}